\documentclass[fleqn,usenatbib]{mnras}

\usepackage{newtxtext,newtxmath}
\usepackage[T1]{fontenc}

\DeclareRobustCommand{\VAN}[3]{#2}
\let\VANthebibliography\thebibliography
\def\thebibliography{\DeclareRobustCommand{\VAN}[3]{##3}\VANthebibliography}

\usepackage{graphicx}	% Including figure files
\usepackage{amsmath}	% Advanced maths commands

\usepackage[usenames,dvipsnames]{xcolor}

\usepackage{soul}
\def\be{\begin{equation}}
\def\ee{\end{equation}}
\def\ba{\begin{equation}\begin{array}}
\def\ea{\end{array}\end{equation}}
\def\bi{\begin{itemize}}
\def\ei{\end{itemize}}
\def\mb{\mathbf}
\title[
Coherent and incoherent structures in fuzzy dark matter halos
]{Coherent and incoherent structures in fuzzy dark matter halos}

\author[I-Kang Liu et al.]{
I-Kang Liu,\thanks{E-mail: i-kang.liu1@newcastle.ac.uk}
Nick P. Proukakis,\thanks{E-mail: nikolaos.proukakis@newcastle.ac.uk}
and Gerasimos Rigopoulos\thanks{E-mail: gerasimos.rigopoulos@newcastle.ac.uk}
\\
School of Mathematics, Statistics and Physics,
Newcastle University, Newcastle upon Tyne, NE1 7RU, United Kingdom\\
}

\date{Accepted XXX. Received YYY; in original form ZZZ. Comments:}
\pubyear{2022}

\begin{document}
\label{firstpage}
\pagerange{\pageref{firstpage}--\pageref{lastpage}}
\maketitle

\begin{abstract}
We show that fuzzy dark matter halos exhibit spatial differentiation in the coherence of the field configuration, ranging from completely coherent in the central solitonic core to incoherent outside it, with a significant drop of the dimensionless phase-space density with increasing radius. The core is a pure condensate, overlapping perfectly with the Penrose-Onsager mode corresponding to the largest eigenvalue of the one-particle density matrix. The virialized outer halo exhibits no clear coherence as a whole upon radial and temporal averaging but can be described as a collection of local, short-lived quasi-condensate lumps, exhibiting suppressed fluctuations, which can be identified with the structures commonly referred to as granules. These localized regions are separated by vortices that form a dynamical web inhibiting phase coherence across the entire halo. We further examine the core oscillations, finding that they are accurately described by two time-dependent parameters characterizing the size of the core, $r_c(t)$, and a crossover region, $r_t(t)$. For the halos in our merger simulations this feature is reflected in the (anti-)correlated oscillation of the peak value of the density power-spectrum. The halo’s turbulent vortex tangle appears to reach a quasi-equilibrium state over probed timescales, with the incompressible component of the kinetic energy exhibiting a characteristic $k^{-3}$ tail in its spectrum, indicative of a $\rho\sim r^2$ density profile around the quantum vortex cores. Comparison of the peak wavenumbers in the corresponding power-spectra shows the inter-vortex spacing and the granule length scale in the outer halo to be very similar and slightly above the core size.
\end{abstract}


%%%%%%%%%%%%%%%%%%%%%%%%%%%%%%%%%%%%%%%%%%%%%%%%%%

%%%%%%%%%%%%%%%%% BODY OF PAPER %%%%%%%%%%%%%%%%%%

\section{Introduction}
The idea that dark matter may be made up of bosonic particles, light enough to exhibit coherent dynamics and wave-like behaviour on galactic scales, 
was introduced some time ago~[\cite{Hu:2000ke}] as a solution to a number of perceived problems exhibited by CDM on sub-galactic length scales [\cite{2015PNAS..11212249W, 2017ARA&A..55..343B, 2017Galax...5...17D}] and has attracted significant attention recently, especially after the pioneering simulations of [\cite{2014NatPh..10..496S}] - see [\cite{Marsh:2016,Hui:2021,2021A&ARv..29....7F}] for recent reviews. Sometimes referred to as Fuzzy Dark Matter~(FDM), Wave Dark Matter or $\psi$-Dark Matter~($\psi$DM), in contrast to the prevailing Cold Dark Matter model~(CDM), it presents a different phenomenology to that of CDM on short scales but is indistinguishable from it on longer scales [\cite{2017PhRvD..95d3541H}] and can thus reproduce CDM's remarkable ability to explain the whole history of structure formation in the universe [\cite{Frenk-White_DM}]. The main parameter delineating these two regimes is the boson's mass with a usually quoted value within an order of magnitude of $m\sim 10^{-22}$ eV/$c^2$ in FDM models; particles of such mass, and with typical velocities found in halos hosting Milky Way-sized galaxies, acquire a de Broglie wavelength of $\mathcal{O}(1)$ kpc and can therefore exhibit wave-like features at galactic scales. Various large scale structure observations have begun placing constraints on the boson's mass [\cite{2016ApJ...818...89S, 2021PhRvL.126g1302R, Dentler:2022}] but a lot of work remains to be done, especially to properly include effects on nonlinear scales which can only be approached via numerical simulations [\cite{2014NatPh..10..496S, 2016PhRvD..94d3513S, 2017MNRAS.471.4559M, Fuzzy_May-Springel}].

There is ongoing debate on whether the aforementioned short scale features of CDM cosmological structures pose a fundamental problem for the model, requiring an alteration of dark matter's fundamental properties, or if it is a matter of proper modeling of complex sub-galactic baryonic physics (gas, SN explosions etc) that feed back on the dark matter short scale distribution~[\cite{2017ARA&A..55..343B}]. Even if the latter view ends up explaining all discrepancies, it is nevertheless interesting to explore the implications of the FDM idea with its different phenomenology acting as an extension of CDM that can be constrained, allowing for example to place limits on the mass (or self-coupling) of scalar Dark Matter particles.\footnote{Here we set the self-coupling to zero and focus on the Schr\"{o}dinger-Poisson system of equations. We will raise this restriction by introducing a quartic self-interaction and study the Gross-Pitaevskii-Poisson system in upcoming work. }   

Beyond observationally motivated considerations, examining the Schr\"{o}dinger equation with the inclusion of the gravitational force is an interesting physics exercise in itself. 
The phenomenon of Bose-Einstein condensation is universal~[\cite{ProukakisUBEC}] and has been studied for a long time in various weakly-interacting laboratory-controlled settings, such as atomic (ultracold quantum matter~[\cite{Anderson1995,Davis1995,Pitaevskii2003,2008bcdg.book.....P}]),
%2008cpcw.book..533C]), 
solid-state/optical (exciton-polariton~[\cite{Kasprzak2006}] and photon~[\cite{Klaers2010}]) systems, alongside studies in superfluid liquid Helium [\cite{Kapitza1938,Allen1938,London1938}], in which the strength of interactions leads to a significantly reduced condensate fraction, even in a pure superfluid [\cite{1956PhRv..104..576P}]. 
Unlike the more strongly interacting helium superfluids, atomic/optical condensates lie in parameter regimes where nearly pure condensates (with a condensate particle fraction approaching $100\%$) can be achieved, and facilitate control of many parameters, including trapping potential and interaction strength, with various ultracold atomic species featuring a strong dipolar component and facilitating a controllable interplay between contact and long-range interactions [\cite{Stuhle2005,Ferrier-Barbut2016}].
%settings relevant for the laboratory~\cite{1999RvMP...71..463D}.
Therefore, examining the features introduced by the long range gravitational interaction may lead to behaviours that could bear similarities to atomic condensates forming in external potential traps~[\cite{ODell2000,Papadopoulos2007}].
Accordingly, insights from the field of cold atom physics may be used to further understand the structure and dynamics of FDM halos.

In this paper we examine the structure and coherence of a single virialised FDM halo: in our simulations, such a halo is formed from the merger of a number of initially coherent soliton spheres placed at random initial positions. After the resulting halo has virialised, the well known central soliton has formed and the Navarro–Frenk–White~(NFW) density profile has been established outside it, we examine the coherence of the field configuration in the spatial domain by measuring (through temporal and generally also angular averaging) the first and second order spatial correlation functions
characterizing phase coherence and density fluctuations respectively. We also directly measure the coherent, Penrose-Onsager mode, corresponding to the single-particle density matrix's eigenmode with the largest eigenvalue, which ends up overlapping with the empirical soliton density profile. Such analysis allows us to clearly distinguish a spatial structure consisting of the inner coherent (solitonic) core, a 
crossover region and a largely incoherent outer halo,
which can be parameterized by two characteristic radii $r_c$ and $r_t$ delineating these regimes. Our simulations also reveal the previously reported core oscillations [\cite{2021PhRvD.103b3508L, 2021PhRvD.103j3019C, 2021ApJ...916...27D}] which we find here to be very well captured by assigning time dependence to $r_c$.

Bose-Einstein condensation corresponds to the existence of macroscopic occupation in a particular state, exhibiting off-diagonal-long-range-order (ODLRO) and implying a two-point correlation function which does not tend to zero as $|\mb{r}-\mb{r}^\prime|\rightarrow\infty$, a feature absent in the normal phase.
In textbook statistical mechanics, three-dimensional (3D) homogeneous, non-interacting systems exhibit Bose-Einstein condensation when the so-called (dimensionless) phase space density ${\cal D} = (\rho / m) \lambda_{\rm dB}^3 > 2.612$~[\cite{2008bcdg.book.....P}] (or more generally ${\cal D} \gtrsim O(1)$). In other words, the emergence of Bose-Einstein condensation arises when the characteristic de Broglie wavelength $\lambda_{\rm dB}$ exceeds the average interparticle distance. In this work, we thus also investigate the value of ${\cal D}$, as a relevant probe of the wave-mechanical nature of the system. 

In a typical 3D harmonically-confined atomic system, the central high-density region is dominated by an inhomogeneous coherent condensate component, with lower-density regions around such a core consisting of incoherent (thermal) particles. 
Moreover, the existence of strong phase fluctuations can also lead to the emergence of a so-called quasi-condensate state [\cite{Popov2001,1997PhRvL..79.3331K}], which features (relatively) suppressed density fluctuations, but sustained phase fluctuations.
Such a state, associated with long-wavelength fluctuations destroying the phase coherence, is particularly pronounced in equilibrium lower-dimensional (quasi-2D, quasi-1D) laboratory settings [\cite{PhysRevLett.87.270402,Petrov_2D,Petrov_3D,Dettmer_2001,PhysRevA.66.013615,2006PhRvA..74e3617P,Cockburn2011,Hadzibabic_2011}] (which facilitate a decoupling between characteristic temperatures for density and phase fluctuations) or close to the critical region, and emerges as a transient state during the formation of Bose-Einstein condensation [\cite{2002PhRvA..66a3603B,ProukakisUBEC}].

Fuzzy dark matter halos show a combination of these features, with all four coherence measures we employ resulting in quantitatively compatible radial profiles (as shown in subsequent figure \ref{fig:3}). 
Firstly, such systems typically satisfy the phase-space density condition (see also subsequent parameter regime discussion), thus justifying a classical field description throughout, as implemented in this work. 
Moreover, such systems exhibit behaviour rather analogous to the harmonically-confined atomic systems: in particular, they consist of a coherent solitonic core surrounded by fluctuating regions of significantly lower density and coherence, mediated by an intermediate region over which the system density fluctuations become gradually enhanced.
The solitonic core is established by the attractive gravitational potential, which plays the role of the trap concentrating the particles. Density and phase fluctuations increase gradually, but significantly, over the immediate surrounding region, with such region further embedded within a fluctuating halo state which, {\em on average}, features no phase, or density, coherence. Nonetheless, closer inspection of such outer halo region reveals a much richer structure featuring the co-existence
of patches of slowly-varying density and variable (but often significant) phase coherence, separated by vortices, which are themselves dynamically propagating (but not dissipating) through the halo, ensuring any local coherence does not extend into the surrounding halo regions.

The coherence of both phase and density in the soliton core region remains intact throughout the post-virialization evolution of the halo. In contrast, the outer halo exhibits both phase and density fluctuations\footnote{Interestingly, our findings appear compatible with the qualitative suggestion of \cite{Guth:2014hsa} for the nature of the equilibrium state of axionic particles with attractive self-interactions, including gravity.}; we explicitly show that they peak at the same characteristic `granular' length scale and are suppressed below it, thus giving rise to a rapidly-fluctuating tangle of quantum vortices intertwined with the granular density distribution, as first suggested in~[\cite{2021JCAP...01..011H}]. We further consolidate this picture through studying the relevant power spectra, the second object of this study. It should be noted that, interestingly, for the halos in our merger numerical experiments, the power spectrum of the density field also carries the signature of the oscillating core
%The coherence of both phase and density in the soliton core region remains intact throughout the post-virialization evolution of the halo. In contrast, the outer halo exhibits both phase and density fluctuations; we explicitly show that they peak at the same characteristic `granular' length scale and are suppressed below it, thus giving rise to a rapidly-fluctuating tangle of quantum vortices intertwined with the granular density distribution, as first suggested in~[\cite{2021JCAP...01..011H}]. We further consolidate this picture through studying the relevant power spectra, the second object of this study. It should be noted that, interestingly, for the halos in our merger numerical experiments, the power spectrum of the density field also carries the signature of the oscillating core.  

Beyond the formation of a stable, density and phase coherent solitonic core in virialized halos, the other main difference of FDM compared to CDM is the way vorticity manifests itself and is what we also focus on in this work. Any initial vorticity present in the early universe is expected to decay like $1/a$ in the matter-dominated era, as long as density perturbations remain small and linear perturbation theory is sufficient for describing the evolution of the density field~[\cite{Durrer2008}]. Hence, a vorticity-free initial condition is a standard consideration in late universe cosmology~[\cite{2018JCAP...09..006J}]. As long as CDM can be modeled by a pressure-less perfect fluid, vorticity will not reappear. As perturbations grow however, CDM enters the so called multi-stream regime, the perfect fluid approximation breaks down and vorticity is generated [\cite{2009PhRvD..80d3504P}].
In the FDM model vorticity also arises during halo formation but, unlike corpuscular CDM, a rotational velocity field can only be accommodated through the appearance of quantized vortices, a well-studied  subject in the field of laboratory condensates. The presence of vortices in the central region of a rotating FDM halo could deform it from spherical symmetry  to a disk-spiral shape~[\cite{Zinner2011}].
However, it has been theoretically found (for Riemann-S ellipsoid solutions) that a vortex could not be stable without a strong enough repulsive self-coupling between FDM particles and a heavy enough bosonic mass such that the core density is/becomes low enough to energetically support the presence of a vortex~[\cite{2012MNRAS.422..135R,2014MPLA...2930002R,2021PhRvD.104b3504D,2021MNRAS.505..802S,2021LTP....47..684N}].
The dynamical instability of a vortex sitting at the centre of a halo leads to its migration out of the core which would temporarily split~[\cite{2021PhRvD.104b3504D}] into a number of fragments dependent on the charge of the vortex~[\cite{2021LTP....47..684N}].
These instabilities lead to the expectation that in a dynamical halo formation no vortices will reside in the core. Indeed, we find that the core region remains shielded from vortices, retaining its density and phase coherence throughout our simulations. 

\begin{figure}%[t]
    \centering
    \includegraphics[width=1\linewidth,keepaspectratio]{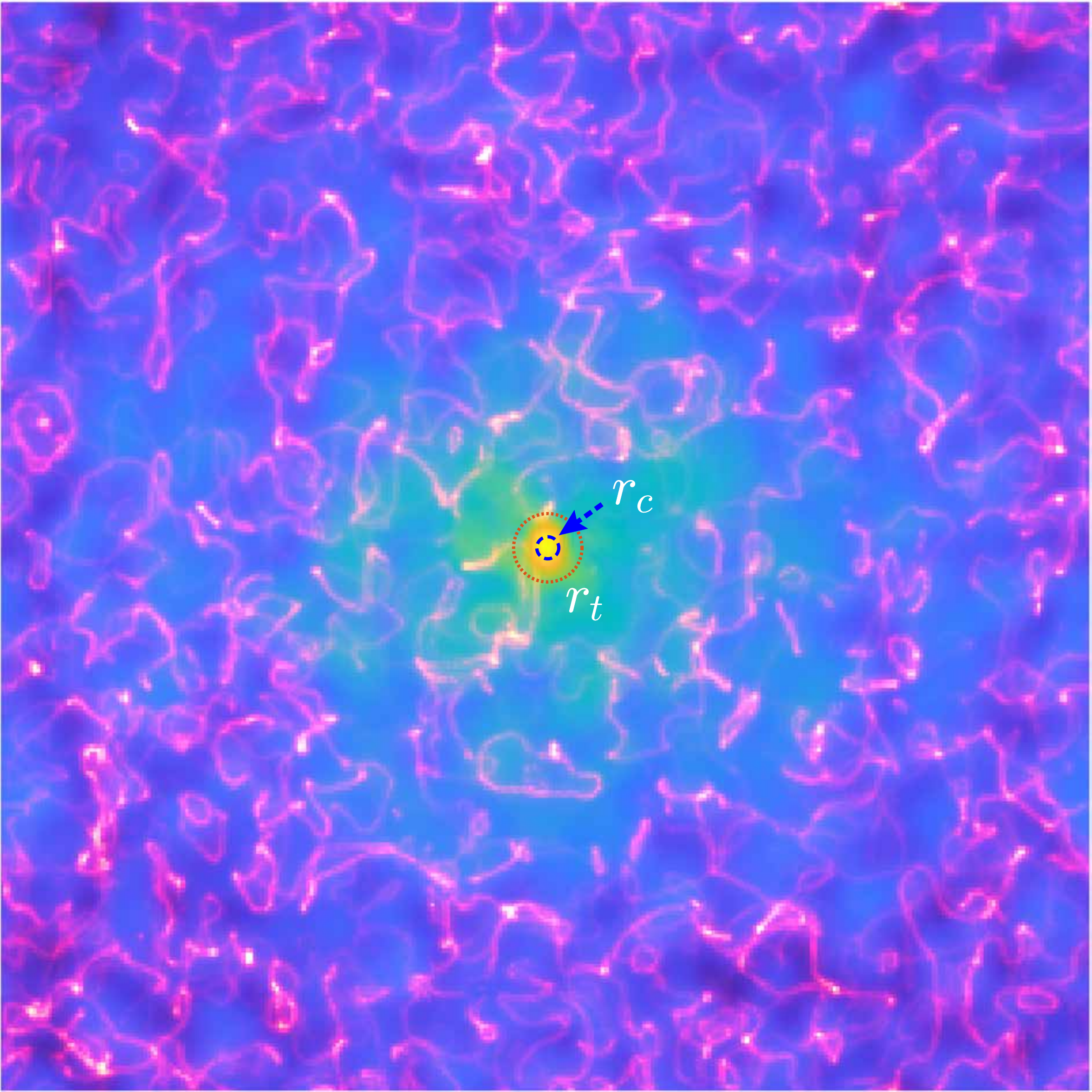}
    \caption{Rendered illustration of the core (inner yellow region) and outer halo (blue region) density profiles, on which the co-existing topologically-closed tangled vortical structures (purple lines) are superimposed, based on our Schr\"{o}dinger-Poisson simulations. The vortical structures, whose characteristic separation scale is a bit larger than that of the soliton core, only appear outside the outer red dashed circle that demarcates a transition zone from coherent to incoherent regions, and do not enter the soliton core region (within the dashed blue circle) once the virialized halo has formed.}
    \label{fig:1}
\end{figure}

As mentioned above, the density beyond the confines of the core is characterised by dynamical granular structures and vortex filaments that form topologically closed loops which deform, stretch and shrink during the time evolution~[\cite{2021JCAP...01..011H}] - see Fig.~\ref{fig:1} and section \ref{sec:vortices}.
The typical inter-vortex separation scale was suggested in~[\cite{2017MNRAS.471.4559M,2021JCAP...01..011H}] to be comparable to the soliton core size. We find it to be a bit larger than that (corresponding to the extended core including the crossover region discussed below) and roughly equal to the granule size. 

Furthermore,~[\cite{2017MNRAS.471.4559M}] has numerically demonstrated that vortices can exhibit turbulent features driven by their reconnections within the vortex tangle that permeates the halo. The existence of a tangle of dynamically-evolving vortices separating regions of near-constant phase -- and thus facilitating short-range superfluid order -- has been previously discussed in the context of condensate formation from an initial highly non-equilibrium state, proceeding through regimes of superfluid turbulence with quasi-condensate local correlations~[\cite{2002PhRvA..66a3603B}]. In such works, which dealt with homogeneous systems, the emerging vortex tangle was numerically observed to relax over long times, with such decaying behaviour also qualitatively persisting in the context of equilibration in an elongated 3D harmonic trap~[\cite{2018CmPhy...1...24L}], where coherence becomes established from the trap centre outwards~[\cite{2020PhRvR...2c3183L}].
In our present setting, we do find the emergence of a phase coherent central solitonic core, surrounded by a halo featuring short-range order across various locations within our simulation box, but with long-range order inhibited by the random motion of the quantum vortices. Unlike such earlier simulations, however, the vortex tangle in the gravitational case does not seem to be noticeably decaying, at least for the rather long timescales probed: such difference could be due to the absence of efficient mode-mixing between different modes of the system in the absence of a self-interaction [\cite{2011JPhB...44k5101C}]. Although the stability of such vortex tangle remains an open question, the present case could constitute a cosmological-scale example of a self-sustained state of strong turbulence which has no current analogue in laboratory condensates.

The incompressible (rotational) kinetic energy spectrum carries a clear signature of the vortices and, through the position of its peak, provides the inter-vortex distance length scale. As mentioned above, we find the latter to be equal to the size of the granular structures that give a ``wavy'' appearance to fuzzy dark matter halos. Beyond the peak, we find the incompressible kinetic energy spectrum in the halo to exhibit a $k^{-3}$ tail, also found in~[\cite{1997PhFl....9.2644N}]. 
This spectrum demonstrates that vortices are not ideal line-like defects but have internal structure and a characteristic size in the superfluid~[\cite{1999RvMP...71..463D}]. Indeed, the $k^{-3}$ tail of the incompressible kinetic energy spectrum implies that $\rho v^2 \approx{\rm const}$ in the vincinity of the vortex centre. Since the velocity profile of a quantized vortex is $v\sim r^{-1}$[\cite{1997PhFl....9.2644N,1999RvMP...71..463D}]
it directly follows that $\rho\approx r^2$ with the centrifugal contribution in the quantum kinetic energy depleting the density around the vortex centre. Using the spectrum of the total velocity field, \cite{2017MNRAS.471.4559M} has verified the velocity profile of a quantized vortex to be close to $v\sim r^{-1}$. On the other hand, \cite{2021JCAP...01..011H} has demonstrated that the density profile exhibits a $\rho\sim r^2$ behaviour in the vicinity of an isolated vortex core in the Schr\"{o}dinger-Poisson case. In our simulations we verify the density profile around FDM vortices in the fully dynamical vortex tangle using the spectrum of the incompressible kinetic energy, a measure more commonly employed in quantum turbulence studies.

%---------------------------------------
%---------------------------------------

A schematic of the core-outer halo features that are at the heart of the study presented in this work is depicted in Fig.~\ref{fig:1}. It is a rendered image for the mass density and spatial distribution of vortices and vortex rings in the final virialized halo. The blue-yellow hues in the figure correspond to the density amplitude while the purple-red is the volume-rendering profile of the modulus square of the velocity field's rotational part in the outer region of the halo. The inner high-density region supports the purely coherent core of condensed axions for $r\leq r_c$ (within the dashed blue circle); this is surrounded by a broader radial ring delineating the crossover region (dotted red circle), where phase and density fluctuations start appearing; at larger radii from the solitonic core the phase coherence is reduced and the system becomes chaotic, where a large number of tangled quantum vortices lie between regions of otherwise nearly-constant density which also exhibit variable local coherence. We will see that the inter-vortex distance and ring size are slightly larger than the extended core diameter $\sim 3r_t$. We discuss the details of this image below, the aim being to characterize these distinct features using a combination of characterization techniques from statistical physics, quantum optics, cold quantum matter, condensed matter and turbulence theory. 

The paper is structured as follows:
In section~\ref{sec:formula}, we discuss the model formulation of FDM and its hydrodynamic mapping and outline our numerical method, modelling scheme and procedure of the performed soliton merger simulation which allows us to characterize the static and dynamical core-halo properties.
In section~\ref{sec:coherence} we discuss the interplay between coherent and incoherent features through various appropriate characterizations, and present the corresponding energy distributions.
In section~\ref{sec:core_oscillation} we discuss the breathing oscillation exhibited in the soliton core and its correlation to the peak wavenumber of the power spectrum for a full halo density profile in the cases we study.
In section~\ref{sec:vortices}, we demonstrate the existence of a large number of vortices in the outer halo and use the granule power and vortex energy spectra to obtain the relation of the granule-size and the inter-vortex distance length scales.
Our conclusions and final remarks are given in section~\ref{sec:conclusion}. Numerical and other technical details, along with some statistical measurements supporting the universality of the presented picture are given in the Appendices.

%%%%%%%%%%%%%%%%%%%%%%%%%%%%%%%%%%%%%%%%%%%%%%%%%%%%%%%%%%%%%%%%%%%
\section{Theoretical background and Numerical Implementation}
\label{sec:formula}
%%%%%%%%%%%%%%%%%%%%%%%%%%%%%%%%%%%%%%%%%%%%%%%%%%%%%%%%%%%%%%%%%%%
\subsection{Schr\"odinger-Poisson equations and corresponding hydrodynamic description}

The ultra-light scalar field making up fuzzy dark matter can be described in the non-relativistic limit by the Schr\"odinger-Poisson system of  equations~(SPE)~[\cite{1969PhRv..187.1767R, 1993ApJ...416L..71W,1994PhRvD..50.3650S, 1996PhRvD..53.2236L, 2003MNRAS.342..176C, 2014PhRvD..90b3517U}]
\be
i\hbar\frac{\partial}{\partial t}\Psi(\mb{r},t)=\left[-\frac{\hbar^2\nabla^2}{2m}+m\Phi(\mb{r},t)\right]\Psi(\mb{r},t)
\ee
where $\Psi=\Psi(\mb{r},t)$ is the wavefunction of fuzzy dark matter in the physical coordinate $\mb{r}$, and the gravitational potential $\Phi(\mb{r},t)$ is determined via the Poisson equation
\be\label{eq:Poisson}
\nabla^2\Phi(\mb{r},t)=4\pi G  \left[\rho(\mb{r},t)-\bar{\rho}\right]
\ee
with $\rho(\mb{r},t)=|\Psi(\mb{r},t)|^2$ the mass density, and $\bar{\rho}$ its spatially averaged, constant value.
As we are focusing on the dynamics of a virialized halo, we would not expect the cosmic expansion to have a strong influence on the halo structures we wish to study here, so we do not account for it in the present work.  
The subtraction of $\bar{\rho}$ from the r.h.s. of (\ref{eq:Poisson}) is reminiscent of the Jeans swindle for a fluid of mass density coupled to Newtonian gravity~[\cite{Binney2011}] but actually has a rigorous mathematical basis for Newtonian gravity~[\cite{Kiessling1999,2011PhRvD..84d3531C}].
It is also associated with the periodic boundary condition for the numerical solution found by the implementation of the pseudo-Fourier spectrum method~[\cite{2008PhRvB..77k5139D}], which is commonly used in most studies and also in this work.

The system involves three main conserved quantities~[\cite{2011PhRvD..84d3531C,2011PhRvD..84d3532C,2021LTP....47..684N,2021MNRAS.505..802S}]: the total mass
\be
M=\int d^3\mb{r} \, \rho(\mb{r},t)
\ee
the energy
\be
E%\left[\Psi^\ast(\mb{r},t),\Psi(\mb{r},t)\right]
=\int d^3\mb{r}\,\,\Psi^\ast(\mb{r},t)\left[
-\frac{\hbar^2\nabla^2}{2m^2}+\frac{1}{2}\Phi(\mb{r},t)
\right]\Psi(\mb{r},t)
\label{eq:energy_functional}
\ee
obtained from the usual expression after integration by parts and ignoring the boundary terms, and the angular momentum 
\begin{equation}
 \mathbf{L}= \int d^3\mb{r} \,\, \Psi^\ast(\mathbf{r},t)\left[\mathbf{r}\times\left(\frac{\hbar}{im}\nabla\right)\right]\Psi(\mathbf{r},t)
\end{equation}
which is zero in the case we examine here (as also confirmed numerically). Expression (\ref{eq:energy_functional}) for the energy functional can also be used as an energy functional of a Bose-Einstein condensate system with the Hartree variational principle to derive the SPE ~[\cite{1999RvMP...71..463D,2011PhRvD..84d3531C,2011PhRvD..84d3532C}].
The source of gravitational potential in the SPE is the collective potential of the mass distribution $\rho(\mathbf{r},t)$ which arises from the long-range, two-body interaction $U_{2B}(\mb{r},\mathbf{r}')=Gm/|\mb{r}-\mathbf{r}'|$ in the context of mean-field description for a Bose-Einstein condensate~[\cite{ODell2000,Papadopoulos2007,2011PhRvD..84d3531C,2011PhRvD..84d3532C,2018PhRvD..98d3509V,Soto:2022}].
The SPE satisfies well-known scaling properties for the spatial scale factor $\Lambda$ ($x \rightarrow \Lambda \hat{x}$)  and boson mass $m \rightarrow \alpha \hat{m}$, see e.g. [\cite{2017MNRAS.471.4559M}], which can be found in Appendix~\ref{appendix:scaling}.

The SPE can be mapped into hydrodynamic equations 
by means of the Madelung transformation,
$\Psi(\mb{r},t)=\sqrt{\rho(\mb{r},t)}e^{i\varphi(\mb{r},t)}$
and 
$\mb{v}(\mb{r},t)=(\hbar/m)\nabla\varphi(\mb{r},t)$. One thus obtains
\be
\frac{\partial}{\partial t}\rho+\nabla\cdot(\rho\mb{v})=0
\ee
and
\be
\frac{\partial}{\partial t}\mb{v}+\frac{\nabla}{m}\left[
\frac{m|\mb{v}|^2}{2}+m\Phi-\frac{\hbar^2}{2m}\frac{\nabla^2\sqrt{\rho}}{\sqrt{\rho}}
\right]=0.
\ee
The equation of motion of $\rho$ is the continuity equation 
while that of $\mb{v}$ contains the gravitational potential and quantum pressure terms. 
The second hydrodynamic equation for the velocity field $\mb{v}$ is in fact identical to the Euler equation for a classical irrotational inviscid fluid in the limit of $\hbar\rightarrow0$.
This indicates that the quantum pressure term can be a source of crucial differences between the classical and quantum fluid but also suggests that classical turbulence can also be present in quantum fluids.
In spite of the SPE and the hydrodynamic equations being mathematically equivalent, the latter cannot fully resolve numerically regions of vanishing density that occur due to destructive interference without further appropriate numerical implementation~[\cite{2018JCAP...09..006J}].
Since sites of vanishing density are places where vortices can form~[\cite{2021PhRvD.103b3508L}], the SPE system is the appropriate framework for studying them.\footnote{It is worth noting that \cite{2011JPhB...44k5101C} has suggested supplementing the hydrodynamic equations by equations of motion for the vortices to handle low density regions of destructive interference.} With the Madelung transformation, the energy of the system, Eq.~(\ref{eq:energy_functional}), can be rewritten as a sum of three different parts,
\be
E=\displaystyle E_\mathrm{ke}+E_\mathrm{qp}+E_{\Phi}= \int d^3£\mb{r}\left[\varepsilon_\mathrm{ke}(\mb{r})+\varepsilon_\mathrm{qp}(\mb{r})+\varepsilon_{\Phi}(\mb{r})\right]
\ee
where $E_\mathrm{ke}$, $E_\mathrm{qp}$ and $E_\Phi$ are classical kinetic, quantum pressure and gravitational potential energies respectively, with the energy densities given by
\be
\varepsilon_\mathrm{ke}(\mb{r})=\frac{1}{2}\rho(\mb{r})|\mb{v}(\mb{r})|^2,\quad \varepsilon_\mathrm{qp}(\mb{r})=\frac{\hbar^2}{2m^2}\left|\nabla\sqrt{\rho(\mb{r})}\right|^2,
\ee
contributed by the Laplacian term in the SPE, and 
\be
\varepsilon_{\Phi}(\mb{r})=\frac{1}{2}\Phi(\mb{r})\rho(\mb{r}).
\ee

Since the velocity field is defined through the gradient of a scalar field, one would expect the system to be irrotational, namely $\nabla\times\mb{v}=0$, which would leave no room for vorticity.
This conclusion only holds however if $\varphi$ is continuously differentiable to second order.
If there are points of vanishing mass density, the phase becomes ill-defined there but the wavefunction can remain single-valued by considering a phase winding around the singularity being $\Delta\varphi=2\pi \kappa$ for integer $\kappa$.
This suggests that a vortical structure in $\Psi$ appears as a vortex line with a divergence in $\mb{v}$ along the vortex core, leading to the tangential velocity being
\be
|\mb{v}(r)|=\frac{\hbar}{m}\frac{\kappa}{r}\,.
\label{eq:vortical_velocity}
\ee
The circulation of the velocity field, specifically for the rotational (incompressible) part over a closed contour on the plane normal to the vortex is
%\be
$\Gamma=\oint_C d\mb{l}\cdot\mb{v}^i=(2\pi\hbar/m)\kappa$
, yielding $v\propto1/r$, where $r$ is the distance from the vortex core, while
the vorticity right on a vortex core at $\mb{r}_q$ is
%\be
$\mb{\sigma}(\mb{r})=\nabla\times\mb{v}(\mb{r})=\kappa\delta(\mb{r}-\mb{r}_q).$
The divergence in $|\mb{v}^i(r\rightarrow0)|$ requires a density depletion at the center of the vortical structure, forming a topological defect known as a quantized vortex.
Such a defect localized around the vortex core is characterized by a length scale $\xi$ which, in the absence of self-interactions, is expected to be of the order of the local de Broglie wavelength~[\cite{2021JCAP...01..011H}].

The existence of vortices in FDM halos was numerically demonstrated in~[\cite{2017MNRAS.471.4559M}] and was further examined in~[\cite{2021JCAP...01..011H}], where it was estimated that there should be 
approximately one vortex line in a unit volume with length scale $\lambda_\mathrm{dB}=2\pi \hbar/m|\mb{v}|$, which should also be the granule length scale. Beyond the velocity field $\mb{v}$ which diverges at a vortex core, a key role in the study of superfluid turbulence is played by the density-associated current 
\be
\mb{F}=\sqrt{\rho}\mb{v}
\ee
which tends to a constant at the vortex core. This is the quantity we will use to study turbulence in the following. A vector field can always be decomposed into compressible (irrotational) and incompressible (rotational) parts via the Helmholtz decomposition, namely,
\be
\mb{v}=\mb{v}^c+\mb{v}^i \textrm{ and }\mb{F}=\mb{F}^c+\mb{F}^i
\ee
satisfying
\be
\nabla\times\mb{v}^c=\nabla\times\mb{F}^c=0\textrm{ and }\nabla\cdot\mb{v}^i=\nabla\cdot\mb{F}^i=0
\ee
and can be computed via Fourier transformation~[\cite{2005JPSJ...74.3248K}].
The rotational component of velocity field is linked to the velocity profile Eq.~(\ref{eq:vortical_velocity}), and the extremely large $|\mb{v}|^i$ localized around the vortex cores can also be used as a probe of vortices.  
As a result, we can further decompose the classical kinetic energy into compressible and incompressible parts,
\be
E_\mathrm{ke}=E_\mathrm{ke}^c+E_\mathrm{ke}^i
\ee
with the corresponding energy densities  $\varepsilon_\mathrm{ke}(\mb{r})=\varepsilon_\mathrm{ke}^c(\mb{r})+\varepsilon_\mathrm{ke}^i(\mb{r})=(1/2)|\mb{F}^c(\mb{r})|^2+(1/2)|\mb{F}^i(\mb{r})|^2$.
Such decomposition, which is standard in quantum fluid studies [\cite{1997PhRvL..78.3896N,1997PhFl....9.2644N,2005JPSJ...74.3248K, 2010PhRvA..81f3630N, 2016PhRvA..94e3632S}], is rather crucial in the context of superfluid turbulence and the study of its arising spectra.
Note that $\mb{F}^i\neq\sqrt{\rho}\mb{v}^i$ and $\mb{F}^c\neq\sqrt{\rho}\mb{v}^c$ essentially contain similar information as the full wavefunction $\Psi$ since the Helmholtz decomposition for $\mb{F}$ also involves the spatial variation of the density.

\subsection{Dimensionless Equations}
To solve the SPE numerically we adopt characteristic reference scales for energy, time and length, respectively
\be
E_\mathrm{ref}=\hbar\sqrt{G\rho_\mathrm{ref}}, \hspace{0.1cm} 
\tau_\mathrm{ref}=\displaystyle\frac{1}{\sqrt{G\rho_\mathrm{ref}}},
\hspace{0.1cm}
\textrm{and}
\hspace{0.1cm}
l_\mathrm{ref}=\displaystyle\left(\frac{\hbar^2}{m^2G\rho_\mathrm{ref}}\right)^{1/4}.
\ee
The dimensionless form of the SPE then reads (we suppress labelling the dimensionless $t$ and $\mathbf{r}$ with different symbols, see Appendix B)
\ba{rl}
\displaystyle i\frac{\partial}{\partial t^\prime}\Psi^\prime(\mathbf{r},t)=&\displaystyle\left[-\frac{\nabla^{\prime2}}{2}+\Phi^\prime(\mathbf{r},t)\right]\Psi^\prime(\mathbf{r}^\prime,t^\prime)
\\\\
\nabla^{\prime2}\Phi^\prime(\mathbf{r},t)=&4\pi\varrho'\left(|\Psi^\prime(\mathbf{r}^\prime,t^\prime)|^2-1\right)%\textrm{ with }
\ea
where $\varrho'=\rho_0/\rho_{\rm ref}$ is introduced to tune the total mass of simulation, $M_{\rm tot}=\rho_0V=(\rho_{\rm ref}l^3_{\rm ref})\varrho' V'$ with the dimensionless volume of the computational box $V'$. We have additionally scaled the density to $\rho_{\rm 0}$ and wavefunction to $1/\sqrt{\rho_{0}}$ giving the dimensionless wavefunction normalized to $V'$. Note that $E_{\rm ref}$ is a very small unit of energy, more appropriate for a single particle. A more useful reference energy unit for the configuration is $\mathcal{E}_\mathrm{ref}\equiv\rho_0 E_\mathrm{ref}l_\mathrm{ref}^3/m = (N E_\mathrm{ref}/V)l_\mathrm{ref}^3$ which refers to the energy contained in a reference volume. These units can be scaled to different physical configurations according to the scaling invariance properties of SPE, Eq.~(\ref{eq:scaling_1a}-\ref{eq:scaling_2b}).
Without any loss of generality we thus present our results in the units of $\mathcal{E}_\mathrm{ref}$, $\tau_\mathrm{ref}$ and $l_\mathrm{ref}$. Further details of our simulation scheme and parameters are given in Appendix~\ref{appendix:numeric}.

%%%%%%%%%%%%%%%%%%%%%%%%%%%%%%%%%%%%%%%%%%%%%%%%%%%%%%%%%%%%%%%%%%%%%%%%%%%%%%

%---------------------------------------
\begin{figure*}%[t]
    \centering
    \includegraphics[width=1\linewidth,keepaspectratio]{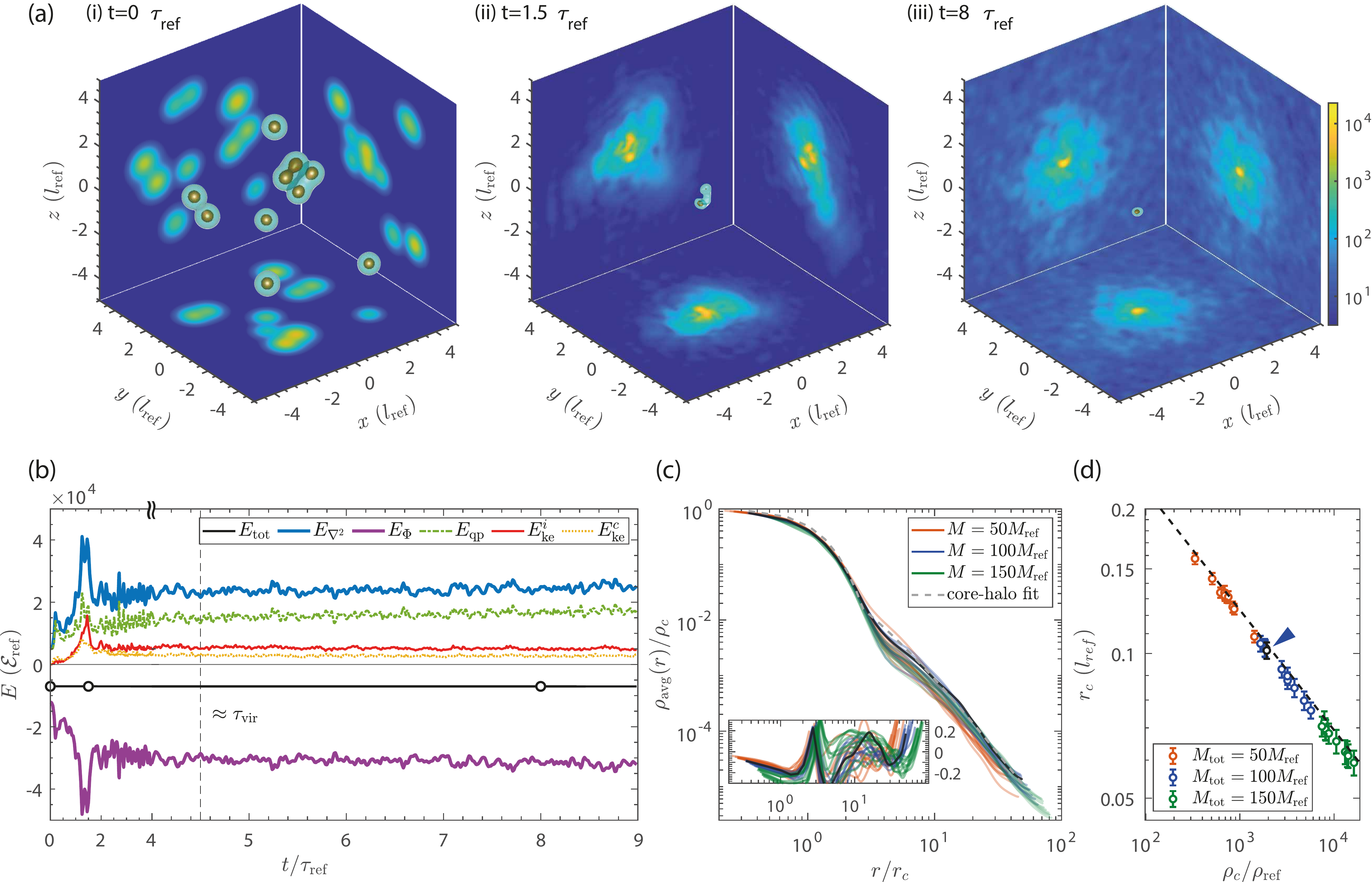}
    \caption{(a) An example of a halo forming in a soliton merger experiment in one of the $M=100M_\mathrm{ref}$ realizations, initiated by 10 randomly distributed solitons of comparable masses. The projection images on each side of the box plot the volume rendering of the density and the yellow and cyan isosurfaces respectively depict the regions bounded within one half and one tenth of the peak density. As is expected in an FDM simulation of such parameters, a soliton core forms at the halo's centre. (b) The energy evolution for the soliton merger simulation shown in (a). The various curves correspond to, in order of appearance in the legend, the total energy ($E_{\rm tot}$, black) which remains constant to better than 1.6\% accuracy, 
    the total kinetic ($E_{\rm ke}$, blue) and gravitational potential ($E_{\rm \Phi}$, purple) energies, the energy associated with the quantum pressure term ($E_{\rm qp}$, green), the incompressible ($E_{\rm ke}^i$, red) and compressible ($E_{\rm ke}^c$, yellow) components of the kinetic energy. Note that $E^{\rm i}_{\rm ke}$, associated with vorticity, is always slightly above $E^{\rm c}_{\rm ke}$, indicating the importance of quantum vortex contribution to the energy budget of the halo. 
    (c) Corresponding virialized radial density profiles, $\rho_\mathrm{avg}(r)$, for 30 different realizations with 3 different total masses and averaged over 390 snapshots, presented in scaled axes (in terms of $\rho_c$ and $r_c$ respectively). The selected data in (a) and (b) are presented in black, and its core-halo fit, Eq.~(\ref{eq:core-halo_fit}), is shown in the dashed grey line. The inset presents the relative error, $[\rho_{\rm avg}-\rho_{\rm cNFW}]/\rho_{\rm avg}$, showing that the core-halo fit can sufficiently capture the core profile with error not exceeding 20\% at any point.
    (d) The core density - radius relation in log-log scale for 30 individual simulations. The error bars here represent the standard deviations for the measured quantities after $t=\tau_\mathrm{virial}$ (see Sec.~\ref{sec:density-r} for more details). The black dashed line plots the core density-radius relation, Eq.~(\ref{eq:soliton_radius}), without any free parameters. The highlighted data point refers to the data presented in (a) and (b) with $E/M_{\rm tot}\approx-70.1 $ and corresponds to the primary dataset analyzed in subplots (a)-(c) and the remainder of this work.
    }
    \label{fig:2}
\end{figure*}
%---------------------------------------

\subsection{Simulation protocol}

We start with a number of randomly distributed idealized solitons with the profile given by Eq.~(\ref{eq:soliton_core}) of comparable masses (determined by a set of uniform random numbers), and allow them to coalesce through their gravitational attraction~[\cite{Schive:2014hza,2016PhRvD..94d3513S, 2017MNRAS.471.4559M, 2022MNRAS.511..943C}] in order to obtain a single virialized halo. This allows us to investigate both (i) the time-averaged spatial and coherence properties of the virialized core-halo system, and (ii) the fluctuating post-virialization dynamics of the system, giving us access to the combined core-halo oscillations and the associated power spectrum.
We simulate the dimensionless SPE in a $V'=L'^3=10^3$ box discretized by $288^3$ grid points imposing a periodic boundary condition from the implementation of Fourier transformation.
Such simulations are repeated 10 times each for 3 different total masses, $50M_\mathrm{ref}$, $100M_\mathrm{ref}$ and $150M_\mathrm{ref}$ ($\varrho^\prime=0.05,\;0.1$ and $0.15$ respectively) with $M_\mathrm{ref}=\rho_\mathrm{ref}(l_\mathrm{ref})^3$, which is about $1.26\times10^6M_\odot$ for our reference parameters (see section \ref{sec:conclusion}). The $50M_\mathrm{ref}$ simulations begin with 5 initial solitons, while 10 initial solitons are chosen for each of the $100M_\mathrm{ref}$ and $150M_\mathrm{ref}$ simulations. The initial solitons are randomly distributed within the simulation box such that for each soliton centre $|x_\mathrm{soliton}|$, $|y_\mathrm{soliton}|$, $|z_\mathrm{soliton}|\leq0.85L^\prime$ and under the constraint that the configuration's centre of mass is close to the centre of our numerical grid. The solitons have very similar masses with a small variance $\sigma_M=0.01/\ln(N_\mathrm{soliton}$).
Due to the slightly different initial conditions, this gives rise to a range of system energies for each total mass configuration, and so we have access to a number of closely-related, yet different, configurations. Specifically, these sets of initial conditions correspond to the system having total energies that lie in the region of $E_\mathrm{tot}/M_\mathrm{tot}\in(-217,-57)\, \mathcal{E}_\mathrm{ref}/M_\mathrm{ref}$.
All simulations proceed to a merger of the initially placed masses, eventually resulting in a virialized halo and exhibiting only a very small energy loss at the level of 1.6\%. This way we can build a complete picture of the physical state and properties of the core-halo system in a general manner through confirmation of our findings over a number of similar mass-energy configurations.

Having analyzed all individual simulations in detail -- and ensured that we observe consistent findings -- the main part of this paper focuses primarily on a single numerical realization shown in Fig.~\ref{fig:2} based on $M=100M_\mathrm{ref}$ and  $E\approx-7006.4\,\mathcal{E}_\mathrm{ref}$, marked by the arrow in Fig.~\ref{fig:2}~(d), with our findings applicable to all conducted simulations. The evolution of 10 initially randomly located solitons is shown in 3 characteristic images, which also depict the volume rendering density profiles, projected to different planes of the simulation box along each perpendicular axis, to give a comprehensive view of the wave-like evolution of density in time. The depicted 3D isosurfaces correspond to a density value equal to one half (yellow), and one tenth (cyan) of the peak value density at a given time, with the former providing an indication of the soliton core size according to the core width definition, Eq.~(\ref{eq:soliton_core}). A movie of this simulation can be accessed via the supplementary material [\href{https://youtu.be/SVLyvERopBw}{SM Movie 1}].

To better appreciate the dynamical virialization process, and following [\cite{2017MNRAS.471.4559M}], we first consider the evolution of the different energy contributions for the presented example in Fig.~\ref{fig:2}~(b).
While the total system energy remains constant (horizontal straight black line), we see significant transfer of energy from gravitational (bottom purple curve) to kinetic (top, blue line) over a typical timescale of $\sim (1-2) \tau_{\rm ref}$, as the initially isolated idealized self-bound solitons approach each other and coalesce due to gravity.
The generated core is heavily excited after the merger, exhibiting a range of dynamical density oscillations, during which heavily reduced, but nonetheless non-negligible, amounts of energy are periodically exchanged between gravitational and kinetic energy, with most exchange concerning the quantum pressure term acting to oppose gravitational attraction.
Gradually, such modulations die out, with the different energy components effectively executing only minor residual noisy oscillations around a converged  final value.
We interpret the time of such occurrence as the time of dynamical equilibration, which we refer to as the virialization time $\tau_\mathrm{vir}$.
For the probed configurations this tends to happen over a time scale $\tau_\mathrm{vir} \sim (4-5) \tau_{\rm ref}$, with the particular depicted example revealing a $\tau_\mathrm{vir} \sim (4.5) \tau_{\rm ref}$ (dashed vertical line in Fig.~\ref{fig:2}(b)). 
In order to accurately perform the temporal averaging discussed in subsequent sections, after $\tau_\mathrm{vir}$ we record successive field configurations every $0.0125\tau_\mathrm{ref}$.

Examining in more detail the evolution of the different kinetic energy components $E_\mathrm{ke}^\mathrm{qp}$, $E_\mathrm{ke}^i$ and $E_\mathrm{ke}^c$, we note that
the quantum pressure energy mostly dominates over the classical kinetic one, and provides support against gravitational collapse on small scales.
In addition, the incompressible part of the classical kinetic energy is always higher than the compressible one, by at least 10\% (with such difference being much more striking at the point of first merging of all input solitons, where it approaches 85\%). The final configuration consists of a single halo containing a quasi-spherical, high-density soliton core at its centre, surrounded by the granular density distribution whose average results in the NFW profile evident at large $r$~[\cite{2021ApJ...916...27D}]. We analyze this further in the sections below.

\subsection{Bimodal Characterization of core-halo profile}

We now discuss the arising density profiles in more detail, using previously documented `bimodal' core-halo profiles. This picture is in a manner reminiscent of condensate-thermal cloud density analysis in confined inhomogeneous ultracold quantum matter.

A dynamically-equilibrated FDM halo contains at its centre what is often referred to as a soliton, characterised by a flat density at small $r$, which can be described by an empirical formula~[\cite{2014NatPh..10..496S,2017MNRAS.471.4559M,2018MNRAS.478.2686C,2021PhRvD.103j3019C}],
\be
\rho_\mathrm{soliton}(r)=\rho_c\left[1+\lambda\left(\frac{r}{r_c}\right)^2\right]^{-8},
\label{eq:soliton_core}
\ee
giving the total mass of the soliton as %{\bf (section 2.5, REF? )}
\be
M_\mathrm{soliton}=\frac{33\pi^2}{1024\lambda^{3/2}}\rho_cr_c^3\approx11.6811\rho_cr_c^3 \,,
\label{eq:soliton_mass}
\ee
where $\rho_c$ and $r_c$ are  mutually dependent parameters - see the following discussion in Sec. 2.5.
The central peak density is $\rho_c$ and the length scale $r_c$ is defined as the radius at which the density drops to half its central value (i.e.~$\rho_c/2)$), giving $\lambda\equiv 2^{1/8}-1\approx0.091$~[\cite{2014NatPh..10..496S}].

The soliton profile, Eq.~(\ref{eq:soliton_core}), has been implemented to probe the solitonic cores of FDM halos, showing good agreement and deviating only by a few percent for $r\lesssim 3r_c$~[\cite{2017MNRAS.471.4559M,2014NatPh..10..496S,2021PhRvD.103j3019C}], which may be taken as approximately the %transition 
crossover radius $r_t$.
The recent theoretical studies ~[\cite{2022MNRAS.511..943C,2022PhRvD.105b3512Y}] and observational surveys~[\cite{2007MNRAS.378...41S,2019MNRAS.490.5451D}] unveil the possibility that the core profile may only extend to a distance less than $3r_c$.
Besides, the mass relation for the core and NFW halo of Schr\"odinger-Poison system is not yet fully determined~[\cite{2022MNRAS.511..943C}].

Even though a lot of attention in the literature has focused on the pure solitonic core, which is indeed a novel feature of FDM compared to the cuspy central region of CDM halos, the whole halo is not just composed of the core but also exhibits the well-known NFW density profile at larger $r$, predicted in CDM and already verified in many FDM simulations. 
Thus, one can approach such a core-outer halo density profile as consisting of the soliton core and NFW profile with a crossover %transition 
radius $r_t$~[\cite{2018PhRvD..97j3523L,2018MNRAS.475.1447B,2022PhRvD.105j3506Z,2022MNRAS.511..943C,2021PhRvD.103j3019C}], defining an inner and outer halo via   
\be
\rho_\mathrm{cNFW}(r)=\Theta(r-r_t)\rho_\mathrm{soliton}(r)+\Theta(r_t-r)\rho_\mathrm{NFW}(r)
\label{eq:core-halo_fit}
\ee
where $\Theta(r)$ is the Heaviside step function, and the NFW profile is given by
\be
\rho_\mathrm{NFW}(r)=\rho_h\left(\frac{r}{r_h}\right)^{-1}\left[1+\left(\frac{r}{r_h}\right)\right]^{-2}
\label{eq:halo}
\ee
featuring the halo length scale $r_h$ and an $r^{-3}$ trend at large $r$.
The continuity of $\rho$ at $r_t$ limits the number of free parameters, determining the value of $\rho_h$ as 
\be
\rho_h=\rho_c\left[1+\lambda\left(\frac{r_t}{r_c}\right)^2\right]^{-8}\left(\frac{r_t}{r_c}\right)\left[1+\left(\frac{r_t}{r_h}\right)\right]^2\,.
\ee
It is worth noting that the continuity of $\rho$ does not necessarily give a smooth differentiable density profile; a smooth transition has been tested in [\cite{2022MNRAS.511..943C}], giving a rather small validity region of the core region, namely, a rather small $r_t$ which could be smaller than $r_c$. 
Besides, the core density can be typically one or even above two orders of magnitude higher than the density at the %transition 
crossover radius~[\cite{2014NatPh..10..496S, 2016PhRvD..94d3513S,2017MNRAS.471.4559M,2018PhRvD..97j3523L,2020PhRvL.124t1301S,2021PhRvD.103b3508L,2021ApJ...916...27D,2022PhRvL.128r1301S,2021PhRvD.103j3019C,2022PhRvD.105j3506Z}].% (AND MORE PAPERS?). 

We now investigate how close the core-halo profile given by Eq.~(\ref{eq:core-halo_fit}) matches the actual radial density distribution in our simulations. To identify the halo's centre we note that the positions of the peak density and potential minimum are nearly identical or only marginally different~[\cite{2020PhRvL.124t1301S}], and thus we associate the halo center to the position of peak density without loss of generality.
We also observe random-like motions in both of these points, with the mean displacement being less than a fraction of the core radius, as already reported in ~[\cite{2021PhRvD.103b3508L,2020PhRvL.124t1301S,2021ApJ...916...27D}].
The actual position of the core's centre is actively traced in the simulation, and, after compensating for this random motion by re-centering the halo, we average the density profiles over 390 snapshots for the presented primary data after $\tau_\mathrm{vir}$ with a time spacing of $0.0125\tau_\mathrm{ref}$ to obtain the time averaged density $\rho_\mathrm{avg}(\mb{r})$.
Such averaged profiles are shown in Fig.~\ref{fig:2}~(c) for all of our simulations in scaled axes, $\rho/\rho_c-r/r_c$.
The radial profile is then obtained by a further angular average 
$f(r)=\sum_{r\leq|\mb{x}_i|<r+\Delta x}f(\mathbf{x}_i)$
with $f(\mb{r})=\rho(\mb{r},t)=|\Psi(\mb{r},t)|^2$. The best fit to Eq.~(\ref{eq:core-halo_fit}) is obtained in log-log scale by examining the least value of the $\chi^2$ error among all possible choices of $r_t\in[1, N_x/2]\Delta x$ with the space resolution $\Delta x$. 
Such a process uniquely identifies the spatial soliton extent ($r_c$), the outer crossover region ($r_t$) and the NFW profile length scale ($r_h$).

Later, in Sec.~\ref{sec:core_oscillation}, we will extend such fitting to the time-domain, making both $r_{c}$ and $r_{t}$ (and thus also $\rho_c$) time-dependent. To our best knowledge the fitting has only been applied to the time-averaged halo profile so far. The fitting for the dynamical halos gives the time-dependent length scales and, as we will discuss below, we also find that the oscillation in the core radius drives an oscillatory power spectrum %, providing a possible direction for observational analyses
 in our halos.

\subsection{Verification of the $\rho_c-r_c$ relation}\label{sec:density-r}

Inspired by [\cite{2011PhRvD..84d3531C}] and following~[\cite{Milos_inpreparation}], the static soliton energy (ignoring the classical kinetic energy part) can be analytically computed as
\ba{rl}
E_\mathrm{soliton}=&E_\mathrm{qp}^\mathrm{soliton}+E_\Phi^\mathrm{soliton}
\\\\
=&\displaystyle\frac{143}{1024}\frac{\hbar^2\pi^2\rho_cr_c}{m^2\lambda^{1/2}}-\frac{152}{96115}\frac{\pi^3G\rho_c^2r_c^5}{\lambda^{5/2}}
\\\\
=&\displaystyle\frac{13}{3}\frac{\hbar^2\lambda}{m^2}\frac{M_\mathrm{soliton}}{r_c^2}-\frac{8506}{2793}\frac{\lambda^{1/2}}{\pi}\frac{GM_\mathrm{soliton}^2}{r_c}~\;.
\ea
Considering the energetically stable state for an isolated soliton, free from fluctuations, and taking $r_c$ as the variational parameter, we set $dE_\mathrm{tot}(r_c)/dr_c=0$ to obtain analytical equilibrium relations connecting the spatial extent of the solitonic core $r_c$ to the soliton mass $M_\mathrm{soliton}$ and peak density  $\rho_c$.
%$M_\mathrm{soliton}-r_c$ and $\rho_c-r_c$ relations for the equilibrium state:
\ba{rl}
r_{c}=&\displaystyle\frac{1107}{389}\frac{\hbar^2\pi\lambda^{1/2}}{m^2GM_\mathrm{soliton}}=\left(\frac{11303\lambda^2}{128}\frac{\hbar^2}{m^2G\pi}\right)^{1/4}\rho_{c}^{-1/4}
\\\\
\approx&\displaystyle23.686\left(\frac{2.5\times10^{-23}\textrm{eV}/\textrm{c}^2}{m}\right)^{1/2}\left(\frac{10^3M_\odot\textrm{kpc}^{-3}}{\rho_c}\right)^{1/4}\textrm{ kpc}.
\label{eq:soliton_radius}
\ea
Such relations are in agreement with many independent numerical simulations~[\cite{2014NatPh..10..496S, 2017MNRAS.471.4559M}], and also satisfy the SPE's scaling invariance ~[\cite{1990PhRvD..42..384S, 2004PhRvD..69l4033G}] (see also Appendix~\ref{appendix:scaling}).
Recalling Eq.~(\ref{eq:soliton_mass}), the mass of a soliton with radius $r_{c}$ can be written as
\ba{rl}
M_{\mathrm{soliton}}=&\displaystyle\frac{1107}{389}\frac{\sqrt{\lambda}\pi\hbar^2}{m^2G}r_{c}^{-1}
\\\\
\approx&\displaystyle 3.68\times10^7\times\left(\frac{2.5\times10^{-23}\textrm{eV}/\textrm{c}^2}{m}\right)^2\left(\frac{r_c}{\textrm{kpc}}\right)^{-1}M_\odot
\label{eq:soliton-mass}
\ea
which is compatible with previous numerical studies.

Having obtained $r_c$ from the fit, we examine the core-radius relation for the soliton, Eq.~(\ref{eq:soliton_radius}). The core density $\rho_c$ after $\tau_\mathrm{vir}$ oscillates by about 50\% around the time-averaged value which we take as the value of $\rho_{c}$. The core radius as function of $\rho_c$ obtained from our simulations is plotted in Fig.~\ref{fig:2}~(d) and the trend  matches the derived relation Eq.~(\ref{eq:soliton_radius}) without any free parameters.

The nice agreement with the expected $\rho_{c}-r_c$ trend indicates that the empirical core profile is a suitable one for the core region of the FDM halo.
There is a small deviation for $r_c<0.08l_\mathrm{ref}\lessapprox3\Delta x$ in $M=150M_\mathrm{ref}$ simulations, suggesting a soliton needs to be at least 3 grid points across to be well resolved. Nevertheless, the overall trend is robust. It is worth emphasising that in our simulations multiple excitations are generated during the merger history of the final halo, and it is consequently hard to define a precise soliton mass. However, the core density and radius relations above can still be taken as useful and relatively accurate guides.

%%%%%%%%%%%%%%%%%%%%%%%%%%%%%%%%%%%%%%%%%%%%%%%%%%%%%%%%%%%%%%%%%%%

\section{Coherent and Incoherent Features of the core-halo system}
\label{sec:coherence}
Having discussed our core-halo system generation protocol and fitting schemes, we next proceed to discuss key features of such virialized systems in terms of spatial coherence, power spectra and dynamics.

\subsection{Condensation and spatial correlation functions}
\label{sec:correlation_fns}

The phenomenon of Bose-Einstein condensation has been extensively studied both theoretically and experimentally in laboratories using ultracold atomic gases (among other systems). 
In most configurations, the system evolves towards (potentially through highly non-equilibrium states, e.g. as those occurring during the condensation process~[\cite{2002PhRvA..66a3603B,ProukakisUBEC,Navon-BeugnonJPBReview,2018CmPhy...1...24L}], and eventually relaxes to, a state of (at least local) thermal equilibrium, described by the associated Bose-Einstein distribution with a fixed temperature and chemical potential, a situation applicable in many experimental setups. However, this assumption cannot be made in the case of a collection of self-gravitating bosons such as those that make up cosmological halos in FDM, because such systems are neither thermalized nor have a well-defined chemical potential in the expanding Universe. Nonetheless, condensation will still occur in this circumstance and it is expected that, like with any collection of bosons, ultralight dark matter will Bose-Einstein condense when the particle number in a volume characterized by the de Broglie wavelength scale is much larger than order unity [\cite{2009PhRvL.103k1301S, Sikivie}]. In cosmological settings, this criterion has already been associated with FDM exhibiting distinctive quantum, or ``wavy'', behaviour~[\cite{2014NatPh..10..496S, 2015PhRvD..92j3510B, Hui:2021}].

A standard criterion delineating the emergence of Bose-Einstein condensation in an equilibrium setting is the
value of the dimensionless phase-space density ${\cal D}$, with the textbook critical value for a non-interacting, homogeneous, 3D condensate being ${\cal D} > \zeta(3/2) \approx 2.612$ [\cite{huang2000statistical,Pitaevskii2003,2008bcdg.book.....P}].
Since in the FDM cosmological context the stationary-states are not thermal (unlike many laboratory condensate systems), one is restricted to using the more general relation $\lambda_{\rm dB} = h / p$ as opposed to the thermal de Broglie wavelength $\lambda_{\rm dB} = \lambda_{\rm T} = \sqrt{2 \pi \hbar^2 / m k_B T}$ used in ultracold superfluid studies.
Obviously, both density and velocity are, in general, functions of ${\bf r}$ in the halo, so here
$p({\bf r}) = m v({\bf r})$. After averaging over both azimuthal directions (to remove angular dependence) and time (during the post-virialized evolution), and under the ergodicity assumption within which temporal averages in steady-state approach the ensemble average~[\cite{2008AdPhy..57..363B,2018CmPhy...1...24L,2020PhRvR...2c3183L}], denoted by $\langle \cdots \rangle$ -- and taking for the sake of our present arguments the simplest known limit of a 3D non-interacting homogeneous equilibrium condensate -- we thus obtain the radially-dependent condensation criterion 
\be\label{cond_crit-1}
{\cal D}(r)=\frac{\rho(r)}{m} \lambda_\mathrm{dB}^3
=\frac{h^3}{m^4} \frac{\langle\rho(r)\rangle}{\langle|{v}(r)|\rangle^3}  \gtrsim 
\zeta(3/2) \approx 2.612\,.
\ee
Furthermore, in terms of the dimensionless density $\rho^\prime(r)$ and velocity ${v}^\prime(r)$ provided by the numerical simulation we can write
\be
{\cal D}(r)
=\frac{h^3\rho_\mathrm{ref}}{m^4v_\mathrm{ref}^3}\frac{\langle\rho^\prime(r)\rangle}{\langle|{v}^\prime(r)|\rangle^3}={\cal D}_\mathrm{ref}\frac{\langle\rho^\prime(r)\rangle}{\langle|{v}^\prime(r)|\rangle^3}~,
\ee
where 
\begin{equation}
{\cal D}_\mathrm{ref}=\displaystyle\frac{h^3\rho_\mathrm{ref}}{m^4v_\mathrm{ref}^3}~\;. 
\end{equation}
As we discuss later, it is useful to consider the quantity ${\cal D}(r)/{\cal D}_{\rm ref}$ which has a universal character that can be applied for different boson masses. This exhibits a significant drop of many orders of magnitude with increasing radius, with its value dropping to $\sim O(1)$ at the crossover radius $r_t$, beyond which point its rate of decrease becomes noticeably reduced. Regardless of the high occupation numbers for small boson masses across our entire simulation grid (see section \ref{sec:conclusion}), there are in fact qualitative differences in the state of the field between the core and the outer halo. Furthermore, the term "Bose-Einstein condensate" typically refers to a state which not only exhibits a high number of particles per de Broglie volume but also a suppression of phase and density fluctuations~
[\cite{huang2000statistical, Pitaevskii2003,2008bcdg.book.....P}]. We therefore introduce below commonly used, and appropriately normalized, spatial correlation functions which characterize the state of the system with respect to such variables. Although, as shown later, the core-halo system oscillates dynamically, once it has virialized such dynamics are perturbations around a steady-state. This makes our time-averaging protocol, used for extracting smoother spatial correlation functions, appropriate for the inner regions. Nonetheless, the surrounding region features intense fluctuations in both the spatial and temporal domain, and we comment further on that below.
%legitimate.

The textbook definition of Bose-Einstein condensation relies on the emergence of Off-Diagonal Long-Range Order (ODLRO)~[\cite{Pitaevskii2003}], i.e.~ the fact that the equal-time one-body density matrix does not vanish at large distances (as $(|{\bf r} - {\bf r'}|) \rightarrow \infty$) but remains finite ~[\cite{1956PhRv..104..576P}]. 
This is best quantified in the normalized first order spatial correlation function\footnote{Note that the general characterization in terms of appropriate field operators, is here replaced by corresponding expressions in terms of classical fields.} 
\be
g_1(\mb{r})=g_1(\mb{0},\mb{r})=\frac{\langle\Psi^\ast(\mb{0})\Psi(\mb{r})\rangle}{\sqrt{\langle|\Psi(\mb{0})|^2\rangle\langle|\Psi(\mb{r})|^2\rangle}}
\ee
quantifying the change of \emph{phase coherence} from the (numerically identified) centre of the solitonic core (${\bf r}={\bf 0}$) to a general radial position ${\bf r}$ away from it.
For a Gaussian fluctuating field, full coherence and complete incoherence are characterized by  $g_1\approx1$ and $0$ respectively (in the limit ${\bf r}  \rightarrow \infty$). We also consider the equal-position second-order correlation function
\be
g_2(\mb{r})=g_2(\mb{r},\mb{r},\mb{r},\mb{r})=\frac{\langle|\Psi(\mb{r})|^4\rangle}{\langle|\Psi(\mb{r})|^2\rangle^2}=\frac{\langle[\rho(\mb{r})]^2\rangle}{\langle\rho(\mb{r})\rangle^2},
\ee
which characterizes \emph{local density fluctuations} in the system, charting the spatial distribution of the degree of local coherence.
Completely suppressed local density fluctuations, consistent with a pure condensate, imply $g_2=1$, while
$g_2=2$ for an incoherent (chaotic) Gaussian field.
In what follows we present the second-order correlation in the form of $2-g_2({\bf r})$ in order to provide a consistent visualization for the two coherence measures.

\begin{figure*}%[t]   
    \centering
    \includegraphics[width=0.925\linewidth,keepaspectratio]{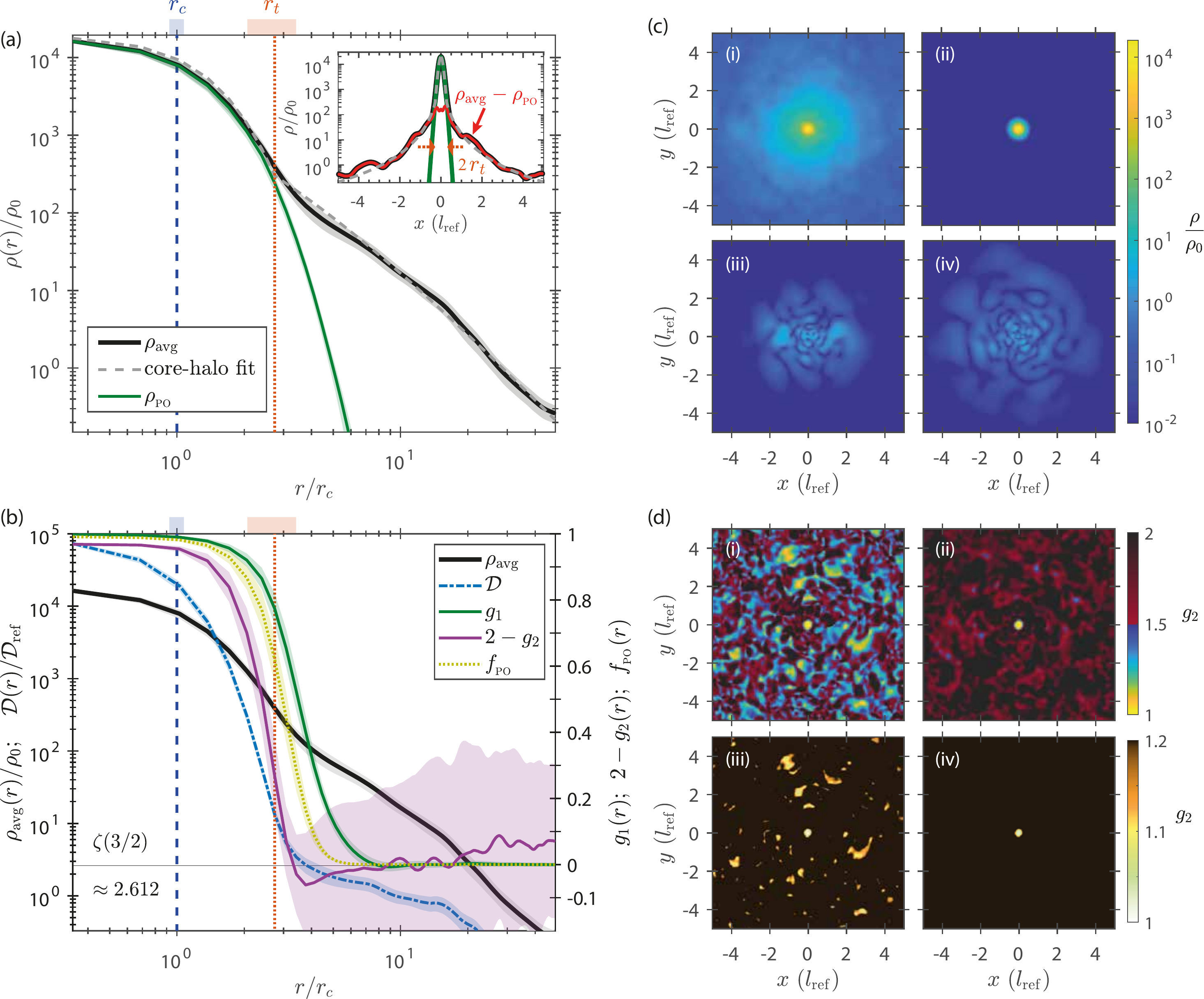}
    \caption{
Characterization of core-halo coherence.
    (a)-(b) Different condensation and coherence measures for the simulation data of Fig~\ref{fig:2}(a), highlighting the crossover from coherent to incoherent field configurations as a function of radius, scaled to numerically-obtained radius $r_c$ of the solitonic core.
    All radial cases show angular- and time-averaged results, taken over $390$ snapshots during the $t\geq\tau_\mathrm{vir}=4.5\tau_\mathrm{ref}$ evolution period in our reference simulation. For comparison, both subplots also show the averaged density profile ($\rho_{\rm avg}$, black line), and locations of fitting parameters $r_c$ (dashed blue vertical line) and $r_t$ (dotted red vertical line) of Fig~\ref{fig:2}(c).
    (a) Penrose-Onsager (PO), or condensate, density ($\rho_{\rm PO} = |\Psi_{\rm PO}|^2$ ) (green)
    and numerical core-halo fit of $\rho_{\rm avg}$ (dashed grey line) as a function of scaled radius $r/r_c$.
    Inset plots scaled densities $(\rho/\rho_0)$ as a function of $x$ for $y=z=0$, depicting the core-halo fit (dashed grey line), and the decompositions of its exact, numerically-obtained, density, $\rho_{\rm avg}$, into the condensate PO mode (central peak, solid green) and the residual lower density $(\rho_{\rm avg} - \rho_{\rm PO})$ (broad red line), clearly highlighting our decomposition into coherent (green) and incoherent (red) contributions to the total density, in qualitative analogy to condensed and thermal features in harmonically-trapped ultracold atomic superfluids.
    The grey band shows the standard deviation of $\rho(r,t)$ over times beyond $\tau_\mathrm{vir}$, and the green %and purple 
    band depicts the standard deviation of the PO 
    density at a given $r$. (b) The radial profiles for the coherence measures: dimensionless reduced phase space density 
    ${\cal D}(r)$, scaled to the reference value ${\cal D}_{\rm ref}$ (light blue, left axis); $g_1(r)$ (green), $2-g_2(r)$ (purple) and condensate fraction $\rho_{_\mathrm{PO}}(r)/\rho_\mathrm{avg}(r)$.
    A horizontal reference line is drawn at ${\cal D}(r)=2.612$ which is the critical phase space density for condensation for a 3D homogeneous non-interacting Bose gas in the thermodynamic limit. Purple band labels the variance of the time-averaged $(2-g_2(r))$, with such band becoming significant
    at distances $r>r_t$ when $g_2$ approaches the value of 2, valid for a Gaussian fluctuating field.
    (c) Planar profiles for $z=0$ of (i) the time-averaged density $\rho_{\rm avg}$, (ii) the PO density corresponding to the largest eigenvalue, (iii)-(iv) the next two dominant eigenmodes of the single-particle density matrix (with eignevalues $N_2 \approx N_3 \approx 0.1 N_{\rm PO}$, obtained by temporal averaging over entire post-virialized evolution. 
    (d) Corresponding $z=0$ planar profiles of $g_2$ averaged over two different timescales and plotted on two different colourbars to highlight [top] range of values $1 \le g_2 \lesssim 2$, and [bottom] regions of significantly suppressed density fluctuations, values $g_2 \lesssim 1.2$.: Left plots [(i), (iii)] correspond to an averaging timescale $t\in[8,8.5]\tau_\mathrm{ref}$,
    broadly consistent with timescale for noticeable vortex evolution, highlighting local regions of significantly suppressed density fluctuations; Right plots [(ii), (iv)] indicate corresponding results when averaged over the entire post-virialization evolution (i.e.~same averaging timescale as used in obtaining plots (a)-(c)).
    }
\label{fig:3}
\end{figure*}

Having specified our distinct measures of coherence, %condensation, 
we now determine them numerically to characterize the state of our representative virialized core-halo system shown in Fig.~\ref{fig:3}(a).
Such analysis, and its confirmation across different $(E,M)$ combinations [see Appendix~\ref{appendix:all_data}]  provide clear evidence of the properties of the obtained  spatially-distinct regions.
The averaged radial dependence of the aforementioned quantities is shown in Fig.~\ref{fig:3} (b) as a function of $r/r_c$, where $r_c$ is extracted from the fit to $\rho_\mathrm{avg}(\mb{r})\equiv\langle\rho(\mb{r})\rangle$, as previously discussed.

Firstly, we note that although the phase space density ${\cal D}(r)$ 
can be very high for typical FDM boson masses,
the reduced phase space density ${\cal D}(r)/{\cal D}_{\rm ref} \sim O(10^5)$ in the solitonic core centre and decreases rather rapidly when moving radially outwards, reaching the critical value $\sim 2.612$ and exhibiting a clear change in its slope around the edge of the crossover region, $r_t$. It then drops below unity in the outer parts of the halo, not far beyond $r_t$. 

Looking at the spatial coherence measures, $g_1$ and $(2-g_2)$, we note that they are almost perfectly flat and $\approx 1$ within the entire inner solitonic core region ($r < r_c$), signifying the near-complete suppression of both density and phase fluctuations, consistent with the solitonic core being a pure condensate. Within the region $r_c \lesssim r \lesssim r_t$ both coherence measures decrease noticeably, and, for $r \gtrsim r_t$, 
the field becomes incoherent ($g_1  \approx 2-g_2 \approx 0$). 
Our numerical results point towards the potential existence of regions locally exhibiting values of $g_2>2$ (marginally), with large error bars in its determination, shifting concrete conclusions beyond our present numerical investigations: while such values would arise in the case of non-Gaussian fluctuations, this could also here be evidence that the averaging time is not enough to completely wash out such overdensities in the slower-evolving outer parts of the halo (see also Appendix~\ref{appendix:all_data}).
Either way, the combination of angular and extensive temporal averaging obscure information about the existence of smaller-scale, localized, partly coherent density patches of variable (non-radially-symmetric) shapes in the outer halo, which are further discussed below -- see also Fig.~\ref{fig:3}(d).

An alternative way to extract information about the presence of the condensate and relate to the mass density  is by directly identifying those parts of the density contributions which exhibit both full phase, and 
density coherence, in direct analogy to the previously identified correlation functions $g_1$ and $g_2$.
In particular, one can directly 
extract from the numerics of the entire field the corresponding condensate mode. This can be done according to the Penrose-Onsager criterion~[\cite{1956PhRv..104..576P,2005PhRvA..72f3608B,2008AdPhy..57..363B}], which identifies the condensate mode $\Psi_\mathrm{PO}(\mb{r})$ as that mode of the single-particle density matrix\footnote{We remind the reader that our wavefunction $\Psi$ normalizes to the system mass, rather than particle number. } 
\be\label{eq:PO}
\varrho(\mb{r},\mb{r}^\prime)\equiv\frac{1}{m}\langle\Psi^\ast(\mb{r})\Psi(\mb{r}^\prime)\rangle
\ee
which has the largest eigenvalue.

Mathematically, the dominant eigenvalues and corresponding eigenfunctions are identified from the solution of
\be
\int {\rm d}{\bf r'} \varrho ({\bf r}, {\bf r'}) \psi_n(\mb{r'}) =  N_{n} \psi_n(\mb{r})
\ee
where the index $n$ labels the $n$-th eigenfunction $\psi_n$ of $\varrho$. Such identification only makes interpretative sense if one, or few, eigenvalues are dominant: in the former case the system is said to exhibit a well-defined Bose-Einstein condensate, with $\psi_0$ the condensate mode, and $N_\mathrm{PO}=N_0$ the condensate particle number.
A system with no (significantly) prevailing eigenvalue, but exhibiting co-existence of few macroscopically occupied eigenmodes is characteristic of a quasi-condensate state, in which there is no overall phase coherence (except locally) due to the competition between dominant eigenmodes (exhibiting different random phases).

As the exact numerical diagonalization of such a large density matrix is numerically impractical, we instead evaluate the largest eigenvalue of the dynamical steady-state system (post-virialization) by replacing the ensemble average in Eq.~(\ref{eq:PO}) by a time average: Strictly speaking this is true for an ergodic system, and this is a standard technique in the field of ultracold quantum matter\footnote{In ultracold atomic systems, the condensate ultimately reaches an equilibrium state over relevant probing timescales, i.e.~before any losses, not typically included, become important and lead to the ultimate solidification of the quantum gas in such systems.}~[\cite{2008AdPhy..57..363B,2018CmPhy...1...24L,2020PhRvR...2c3183L,Proukakis13Quantum}]. The temporal duration of such averaging  should be chosen to be long enough to wash out random fluctuations, but short enough not to obscure any key underlying physical dynamics being probed (see, e.g.~[\cite{2018CmPhy...1...24L}]).
Post virialization, the solitonic core acquires a near-equilibrium profile (modulo small oscillations - see below), and so the exact averaging duration beyond $\tau_{\rm vir}$ has practically no impact on its properties, provided enough samples are averaged over.
However, more attention is needed when addressing the halo regions, where the quantum vortices and granular structures propagate dynamically; in this case our findings become sensitive on whether the averaging timescale is shorter than, or longer than the typical evolution timescales of such structures, as discussed below.
In order to obtain most averaged results presented in this work, we typically probe
%In the present context this amounts to averaging 
the entire post-virialization timescale of $4.875 \tau_{\rm ref}$.
%averaging over 390 snapshots taken $0.0125\tau_\mathrm{ref}$ apart. 
However, when probing short-time local density correlations [left panels in Fig.~\ref{fig:3}(d)], we also consider a shorter averaging time $t_\mathrm{aver}=0.5\tau_\mathrm{ref}$.
In practice we average over a certain number of snapshots taken $0.0125\tau_\mathrm{ref}$ apart, with respective numbers being 390 (long-time) and 40 (short-time) snaphsots
- see Appendix~\ref{appendix:numeric} for further details.

Our simulations clearly reveal the existence of one macroscopically-occupied eigenmode -- found to have an eigenvalue which is $\approx 10$ times larger than that of any other modes -- followed by a range of few approximately equally important eigenmodes (e.g.~for $n=1$ to $5$ we found that $N_n/N_0\approx0.1$ -- see Appendix~\ref{appendix:data} for details). The three first eigenmodes are shown in Fig.~\ref{fig:3}(c)(ii)-(iv), alongside the corresponding time-averaged density [Fig.~\ref{fig:3}(c)(i)]. The largest eigenvalue mode, or Penrose-Onsager (PO) mode [Fig.~\ref{fig:3}(c)(ii)] clearly coincides with the central soliton (including the crossover region); contrary to this, as expected, the next two eigenmodes have $N_2 \approx N_3 \approx 0.1 N_0$ and a much broader radial extent.
The radially-averaged PO mode is shown by the green line in Fig.~\ref{fig:3}(a), and
corresponds to that part of the density with suppressed density and phase fluctuations.

At this point, it is perhaps instructive to draw a closer analogy between our system and ultracold atomic gases. In the latter system, under a harmonic confining potential, one finds a dominant central coherent profile, i.e.~a condensate (in the shape of an inverted parabola) embedded within an incoherent Gaussian thermal cloud distribution (which is used to identify the system temperature),
thus giving rise to a bimodal configuration\footnote{Note that despite the presence of (typically) repulsive interactions, careful modelling in terms of a condensate and a thermal cloud (see, e.g.~\cite{proukakis2008finite}) reveals that a small (approximately constant density) incoherent component persists even within the condensate profile.}. This is highly analogous to the spatial distribution obtained in our present cosmological simulations. To make this `two subsystems' analogy clearer, we impose here a separation of the numerically-obtained total system density into a condensed (PO mode) and an uncondensed component, obtained by subtracting the PO mode from the average density; these are shown respectively by the green and red lines in the inset to Fig.~\ref{fig:3}(a). 
We stress that such determination is based only on the numerically-obtained density and does not in any way invoke the soliton and NFW profiles.
To highlight the excellent agreement of our PO-based analysis for the density components to the bimodal fit based on Eq.~(\ref{eq:core-halo_fit}), which uses a solitonic core profile in the inner region ($r<r_t$) and an NFW profile for $r > r_t$, we overlay in the inset the total density obtained via Eq.~(\ref{eq:core-halo_fit}) by a grey dashed line.
Such near-perfect agreement implies that the well-studied solitonic core of FDM is a \emph{true inhomogeneous condensate} embedded within a dynamical, and on average incoherent, NFW halo. The condensate fraction is defined by $f_{_\mathrm{PO}}(r)\equiv\rho_{_\mathrm{PO}}(r)/\rho(r)$, showing within $r<r_c$ the halo density is nearly a pure condensate with $f_{_\mathrm{PO}}(r)\sim1$ in agreement with the above coherence measurements. 

Having established that the core is a proper BEC with suppressed density and phase fluctuations, the question naturally arises to what extent any such features can be discerned in the outer halo. A useful quantity characterising an `intermediate' state exhibiting suppressed density fluctuations but no true ODLRO, which is actively discussed in the ultracold atomic community, is the quasi-condensate density.
%which characterizes the degree of suppression of spatial density fluctuations.
This is closely related to $g_2$ and is defined according to~[\cite{PhysRevLett.87.270402,2002PhRvA..66a3603B,2006PhRvA..74e3617P,2008AdPhy..57..363B,Cockburn2011,2020PhRvR...2c3183L}]%Check Blair's citing%[Svistunov 2002?, Blakie 2007?, Cockburn-Proukakis-Henkel-2012] (and)~\cite{2006PhRvA..74e3617P,2020PhRvR...2c3183L},
\be\label{eq:QC}
\rho_\mathrm{qc}(\mathbf{r})=\sqrt{2\langle|\Psi(\mb{r})|^2\rangle^2-\langle|\Psi(\mb{r})|^4\rangle}=\rho_\mathrm{avg}(\mb{r})\sqrt{2-g_2(\mb{r})} \;.
\ee 
As we have already seen in Fig.~\ref{fig:3}(b), the value of $g_2$ extracted as an average over the entire post-virialized evolution can still (marginally) exceed the value of 2, while exhibiting large fluctuations -- a finding consistent across all our simulations (details in Appendix~\ref{appendix:all_data}).
As a result, a radial quasi-condensate density can only be defined within our numerics up to $r \lesssim r_t$, where it practically coincides with the PO mode.
Features beyond such point are obscured by the temporal and angular averaging used to obtain the value of $g_2$ in Fig.~\ref{fig:3}~(b).
%which can conceal spatially- and temporally-localized features in the halo.
Therefore, it is very instructive to consider the full spatial dependence of $g_2({\bf r})$, which will allow us to identify spatially- and temporally-localized features relating  to structures in the halo often referred to in the literature as `granules' (e.g.~[\cite{2014NatPh..10..496S}]).

To visualize this, we look at slices of $g_2$ in the $x$-$y$ plane, taken at $z=0$ [Fig.~\ref{fig:3}(d)] and consider two different time averages (left/right columns) to examine any dynamically evolving features which are washed away by long-time averaging. Moreover, in Fig.~\ref{fig:3}(d) we plot the same information on two different colourbars [top and bottom] to more clearly highlight the key arising features.
Looking first at the same (long) time averaging used to extract $\rho_{\rm avg}(r)$, $\rho_{\rm PO}(r)$ and $g_2(r)$ in Fig.~\ref{fig:3}(a)-(c), we see from Fig.~\ref{fig:3}(d)(ii) the existence of the prominent solitonic core with completely suppressed density fluctuations $g_2 \approx 1$ (yellow core). This is surrounded by regions of intense density fluctuations ($g_2 \approx 2$, dark red regions) which dominate the picture, with only a very small number of regions exhibiting suppressed density fluctuations ($g_2 \lesssim 1.5$, light blue regions).
Such rich structure explains why the quasi-condensate cannot be numerically identified (with our current averaging scheme at least) in the radial direction beyond $r_t$. Over the extended time-averaging, there are in fact no regions, other than the extended solitonic core, with $g_2 < 1.2$ [Fig.~\ref{fig:3}(d)(iv)].

However, such picture changes drastically when considering shorter time averages of 0.5$\tau_{\rm ref}$ [Fig.~\ref{fig:3}(d), left column]: In this case, we see clear evidence of widespread regions of locally (partially) suppressed density fluctuations in the halo [yellow/green regions in Fig.~\ref{fig:3}(d)(i)].
%closer inspection [Fig.~\ref{fig:3}(d)(iii)] reveals that
Although density fluctuations in such regions (where the radially-integrated density is well-described by the NFW profile) are evidently locally suppressed ($g_2 < 1.2$), they are nonetheless not as suppressed as in the core itself ($g_2 \approx 1$) [Fig.~\ref{fig:3}(d)(iii)]. We have further found that, at a given time within the probed interval, the phase over a region of suppressed density fluctuations remains {\em largely} constant. This is a strong  indication of the {\em local} existence of quasi-condensate lumps in the halo, with such lumps destroyed on longer timescales through the chaotic passage of quantum vortices.
We stress that consolidation of the above quasi-condensate discussion would require extensive further analysis, which we defer to future work.

\begin{figure}%[t]
    \centering
    \includegraphics[width=1\linewidth,keepaspectratio]{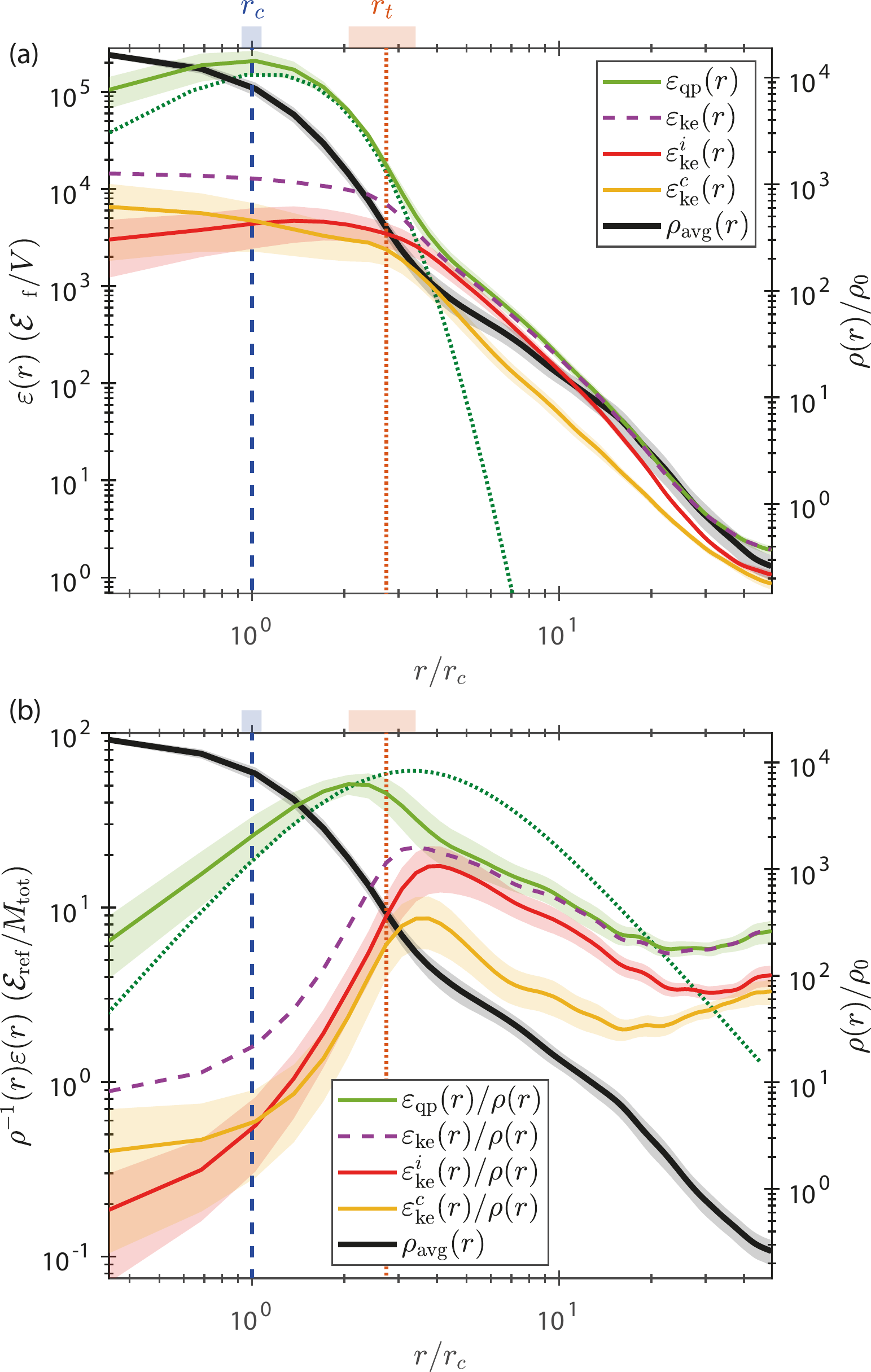}
    \caption{
    (a) The energy distribution as a function of radius for the quantum pressure energy, $\varepsilon_\mathrm{qp}$  (green), and classical kinetic energy $\varepsilon_\mathrm{ke}$ (dashed purple), decomposed into incompressible $\varepsilon_\mathrm{ke}^i$ (red) and compressible $\varepsilon_\mathrm{ke}^c$ (orange) components with $\rho_\mathrm{avg}$ as a reference for the data presented in Fig.~\ref{fig:3}. 
    (b) The radial energy density profile for the incompressible and compressible parts of the kinetic energy per density, a measure of the velocity.
    The bands represent the uncertainty due to the variance in time of the radial profiles for $t\geq\tau_\mathrm{vir}$.
    Dotted green lines show the corresponding prediction for $\varepsilon_\mathrm{qp}$ based on (\ref{eq:Qp-predict})
    }
    \label{fig:4}
\end{figure}

%%%%%%%%%%%%%%%%%%%%%%%%%%%%%%%%%%%%%%%%%%%%%%%%%%%%%%%%
\subsection{Energy density distribution}
\label{sec:energy_distribution}
Next, we consider the spatially-dependent contributions of the different (decomposed) energy densities.
According to Eq.~(\ref{eq:soliton_core}), the quantum energy density of an isolated soliton core can be calculated to be 
\be\label{eq:Qp-predict}
\varepsilon_\mathrm{qp}^\mathrm{soliton}(r)=32\frac{\hbar^2\lambda^2r^2}{m^2r_c^4}\left[1+\lambda\left(\frac{r}{r_c}\right)^2\right]^{-2}\rho_\mathrm{soliton}(r)
\ee
suggesting a quadratic increase of the energy associated to the quantum pressure term in the flat core region.

In Fig.~\ref{fig:4}~(a), we present the distribution  of the classical kinetic energy, broken down into incompressible and compressible parts, as well as the energy associated with the quantum pressure, as a function of radius for the example halo described in this work.
One can readily see the clear dominance of quantum pressure in the core region while the classical kinetic components are at least an order of magnitude smaller (but not zero).
The trend matches well with the estimation from the empirical soliton form up to $r\approx r_t$ where the density transits to the NFW profile for the outer halo region. However, the empirical formula tends to underestimate the quantum pressure contribution, indicating the slightly more dynamical nature of the soliton embedded in a halo as oppossed to an isolated one.
In the outer halo, we find that both quantum pressure and classical kinetic energies are comparable with ${\varepsilon_\mathrm{qp}/{\varepsilon}_\mathrm{ke}}\approx1$ ~[\cite{2017MNRAS.471.4559M}].
As can be clearly seen in Fig.~\ref{fig:4}~(a) and (b), in the outer halo region the incompressible energy contributes slightly more than the classical kinetic energy.

By investigating the energy distribution divided by the mass density, $\rho^{-1}\varepsilon$ in Fig.~\ref{fig:4}~(b), which is essentially the modulus square of effective velocity amplitude per particle, we also find that, as expected, the quantum pressure remains dominant in the core region while the classical kinetic velocities are much smaller. The incompressible part of the classical kinetic energy component is smaller than the compressible part for $r<r_c$; notably $\rho^{-1}\varepsilon^i$ is even less than order unity. As there are no vortices within the core region (see below), its small but non-zero value suggests that the whole core participates in a slow rotational motion relative to points outside it. The compressible part can be larger and is due to the non-vortical motion of the core itself or excitations propagating through it. Visual inspection of the core region in our simulations supports this interpretation: as mentioned above, the core exhibits a random-like walk confined in a small region around the centre and is itself pulsating to a small degree on top of the underlying solitonic density profile. For $r>r_t$ the classical kinetic part becomes comparable to the quantum pressure one, apparently satisfying equipartition. 

%---------------------------------------
\begin{figure*}%[!ht]
    \centering
    \includegraphics[width=1\linewidth,keepaspectratio]{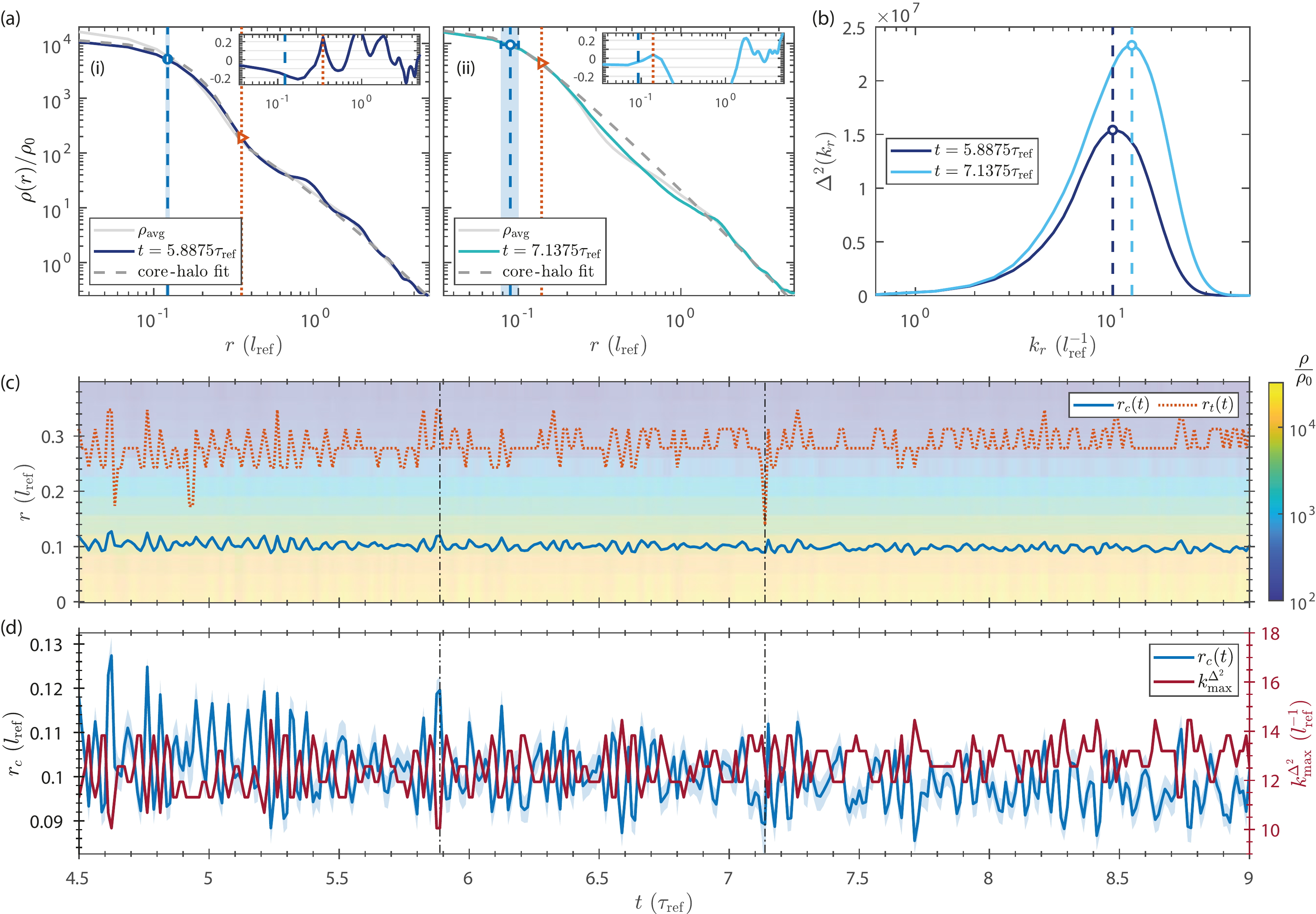}
    \caption{(a) Two dynamical radial density profiles with corresponding core-halo fits for $t=5.8875\tau_\mathrm{ref}$ and $7.1375\tau_\mathrm{ref}$, selected from the temporal evolution of $r_c(t)$ and $r_t(t)$ shown in (c), such that they correspond to the extrema of the $r_t$ oscillation. The blue and red vertical dashed and dotted lines mark the length scales of $r_c(t)$ and $r_t(t)$ with the circle and triangle marking those points on the density profile. %The bands depict the standard deviation of the density for a given radius, showing large fluctuations with order comparable to $\rho(t)$ for $r>r_t(t)$. 
    The insets show the relative error, $[\rho(r,t)-\rho_\mathrm{fit}(r)]/\rho(r,t)$, with the core-halo fit sufficiently capturing the overall profile particularly for $r<r_c(t)$ with less than 20\% error. 
    (b) The corresponding power spectra for the configurations in (a) show a displacement of the peak location of the power spectrum, marked by the vertical dashed lines with hollow circles for the two times $t\geq\tau_\mathrm{vir}$. (c) The temporal evolution of $r_c(t)$ and $r_t(t)$ extracted from the dynamical density profiles - the times shown in (a) are marked by the black dashed-dotted lines and correspond to two extreme variations. The shaded background shows the subtle density deformation in time. (d) The evolution of $r_c(t)$ and the peak momentum of the power spectrum $k_\mathrm{max}^{\Delta^2}(t)$ for the same time windows in (c). A clear anti-correlation for these two quantities can be observed.
    }
    \label{fig:5}
\end{figure*}
%---------------------------------------

\section{Core and power spectrum Oscillations}
\label{sec:core_oscillation}
As we have already seen in Sec.~\ref{sec:density-r}, the (static) central core density and the (static) soliton core width $r_c$ are related. The oscillation of the core density and its frequency have been reported in previous studies~[\cite{2004PhRvD..69l4033G,2018PhRvD..98d3509V,2021PhRvD.103b3508L,2020PhRvL.124t1301S,2021PhRvD.103j3019C,2022PhRvL.128r1301S}].
We find the dominant oscillation frequency, inferred by the Fourier spectra of both $\rho_c(t)$ and $r_c(t)$, to be 
\be
\omega\propto\rho_{c}^{1/2}
\ee
agreeing with the literature and an analytical prediction based on the Gaussian density packet approach~[\cite{2011PhRvD..84d3531C}].
These density oscillations may be driven by interactions between the soliton core and the outer halo~[\cite{2021PhRvD.103b3508L,2022PhRvD.105j3506Z}], resulting in fluctuating anisotropy of the central density~[\cite{2021ApJ...916...27D}].
This relation can also be seen from the scaling invariant relation, Eq.~(\ref{eq:scaling_1a}).

In addition, the Fourier spectra in both $r_c(t)$ and $\rho_c(t)$ (and also the peak wavenumber of the overdensity power spectrum $k_\mathrm{max}(t)$ - see below) exhibit multiple peak-frequency features ~[\cite{2021PhRvD.103b3508L,2021ApJ...916...27D}].
This suggests that there are multiple excitations created in the merger simulation with the nonlinearity in the equations of motion resulting in mode mixing~[\cite{1998PhRvA..57.3818M,2021PhRvD.103b3508L,2022PhRvD.105b3512Y,2022PhRvD.105j3506Z}].
All these lead to fluctuations in the precise core oscillation frequencies, a situation that could also be generic for actual halos, coupled to the baryonic matter. We will report more details about the topic of oscillations elsewhere~[\cite{Milos_inpreparation}].

Going beyond the already studied oscillation of the central density, we now discuss how the core-halo fitting formula (\ref{eq:core-halo_fit}) does not just capture the time-averaged density profile but can also model the dynamical one fairly well.
In Fig~\ref{fig:5}~(a)~(i) and (ii) we plot two radial density profiles with their core-halo fits at two different times, chosen at the extrema of the post-virialized oscillations to demonstrate the broad applicability of our fits.
The core-halo fit successfully captures the overall profile, particularly around the "neck" at $r\approx r_t$, as illustrated in Fig.~\ref{fig:5}~(a)~(i).
Very occasionally in our simulation the extracted value of $r_t$ approaches that of $r_c$, with such an instance shown in Fig.~(a)~(ii) in which the typical head-neck-shoulder profile is not clear.
In Fig.~\ref{fig:5}~(c), we show an example of the evolution of $r_c$ and $r_t$ in time with the two instances depicted in Fig.~(a)~(i) and (ii) marked by the two vertical dashed-dotted lines. Clearly, $r_c$ exhibits far less fluctuations than $r_t$.  

We also examine the dimensionless power spectrum of the overdensity in the field configuration, written as
\ba{rl}
\Delta^2(k_r,t)=&\displaystyle\frac{k_r^3}{(2\pi^2)4\pi k_r^2}\int d\mathbf{\Omega}_kP(\mathbf{k},t)
\\\\
\approx&\displaystyle \frac{k_r^3}{2\pi^2\mathcal{N}_{k_r}}\sum_{k_r\leq|\mathbf{k}|<k_r+\Delta k}P(\mathbf{k},t)\,.
\ea
Here $P(\mathbf{k},t)=\left|\tilde{\eta}(\mathbf{k},t)\right|^2$ is the power with the Fourier transformation of the scaled density $\eta(\mathbf{r},t)=[\rho(\mathbf{r},t)-\rho_0]/\rho_0$ with respect to the spatially averaged density $\rho_0$ which is a constant in time; $\mathbf{\Omega}_k$ the solid angle in the momentum space and  $\mathcal{N}_{k_r}$ is the number of momentum grids within $k_r\leq|\mathbf{k}|<k_r+\Delta k$ and $\Delta k_r=2\pi/L\approx0.63l_\mathrm{ref}^{-1}$.

The power spectrum is time dependent and always exhibits a clear peak, the height and position of which oscillates. Two instances of the power spectrum are shown in Fig.~\ref{fig:5}~(b), corresponding to the profiles shown in (a-i) and (a-ii) of the same figure and also indicate the range over which the power spectrum varies in the virialized halo of our chosen example - at any given instant the power spectrum lies between these two curves. Note that when the peak is located at the maximum wavenumber the core and the crossover region are at their smallest extent, while the minimum peak wavenumber corresponds to the most extended phase of the core during its oscillation.
This correspondence between the peak wavenumber of the power spectrum and the oscillation phase of the core is evident in Fig.~\ref{fig:5}~(d) where we plot both $r_c(t)$ and the peak wavenumber of the power spectrum. It is clear from that figure that the two quantities are almost perfectly anti-correlated. The natural question to ask is whether this power spectrum oscillation is driven by the core alone or whether the granules oscillate too. The dynamical state of the core with its oscillatory and random walk motion has prevented us from subtracting it cleanly from the halo to study the granule spectrum alone, but preliminary results, to be reported elsewhere [\cite{Milos_inpreparation}], indicate that at least most of the oscillatory contribution to the power spectrum comes from the core and the granule spectrum is much less time dependent. Thus, the core's time dependent size shows up in the power spectrum, at least for the halos in the mass and energy regime formed in our simulations. 

%%%%%%%%%%%%%%%%%%%%%%%%%%%%%%%%%%%%%%%%%%%%%%%%%%%%
\section{Vortices and Granules}
\label{sec:vortices}
%---------------------------------------
\begin{figure*}%[!ht]
    \centering
    \includegraphics[width=.9\linewidth,keepaspectratio]{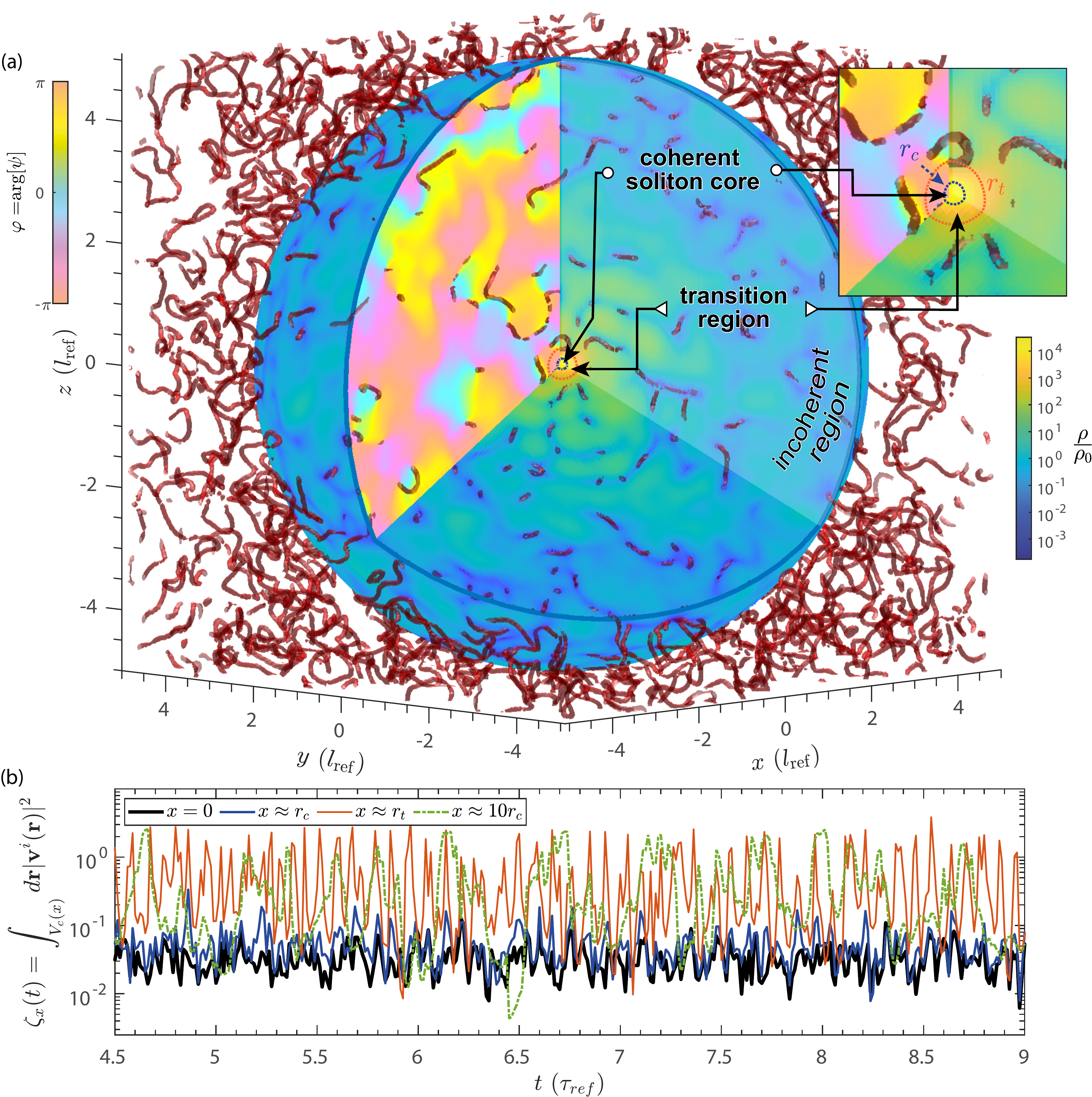}
    \caption{
   % \np{JUST CHECKING: IS THIS MEANT TO BE SCALED TO $\rho_{\rm avg}$???}
    (a) A composite visualization of the halo at $t=8\tau_\mathrm{ref}$, with peak density at the centre of the computational box.
    The red filaments are the isosurfaces of $|\mathbf{v}_{ke}^i|^2=800 l_\mathrm{ref}^2/\tau_\mathrm{ref}^2\approx(\hbar^2/2m^2
    \Delta x^2)$, representing vortex structures in closed loop topologies.
    The slices on the right and bottom of the sliced sphere show the density profile $\rho(\mb{r})/\rho_0$ %\np{[GARY: I AM STILL CONFUSED. FIGURE COLOURBAR IS SCALED TO $\rho_0$ BUT CAPTION NOW SAYS $\bar{\rho}$]}
    on them with colours ranging from blue (low density) to yellow (high density). The phase profile of the wavefunction is presented on the left slice with a pale pink-purple-blue-green-yellow cyclic colour scheme for phases $-\pi\rightarrow\pi$.
    The plot also highlights the characteristic radii $r_c$ (dashed blue) and $r_t$ (dotted red), demonstrating the absence of vorticity within such radial regions -- a feature which persists during the entire probed evolution (see supplementary movie [\href{https://youtu.be/KXCFshb6-90}{SM Movie 2}]
    (b) The evolution of the local integral of the modulus square of the incompressible velocity field, $\zeta(x,t)$, for $x=0$ (black), $x=r_c$ (blue), $x=r_t$ (red), and $x=10r_c$ (dashed green) for $r_c$ computed from the time averaged density profile shown in Fig.~\ref{fig:3}.
    }
    \label{fig:6}
\end{figure*}
%---------------------------------------

\subsection{Vortical structures outside the core}

The energy distributions illustrated in Fig.~\ref{fig:4} show that, opposite to the core region where the quantum pressure dominates, in the outer halo region, or more specifically for $r>r_t$, the classical kinetic energy is as important as the quantum pressure one, namely $\varepsilon_\mathrm{ke}(r)\approx\varepsilon_\mathrm{qp}(r)$.
In addition, the contribution of the kinetic energy's incompressible component is slightly higher than that of the compressible component, as discussed in Sec.~\ref{sec:energy_distribution}.
The rotational/incompressible part of the velocity $\mathbf{v}^i$ and density current $\mb{F}^i$ in a scalar field can only receive contributions from quantized vortices. 
In the vortex core the velocity exhibits a divergence, see Eq.~(\ref{eq:vortical_velocity}), while the density current converges to a constant. Here we use the high incompressible velocity amplitude as a probe in order to visualize vortex features in the outer halo [\cite{2021JCAP...01..011H}]. The vortices identified are plotted in Fig.~\ref{fig:6}~(a) where the isosurface of $|\mb{v}^i(\mb{r})|^2=800l_
\mathrm{ref}^2/\tau_\mathrm{ref}^2\approx(\hbar^2/2m^2)(\Delta x)^{-2}$ is illustrated in red, vividly showcasing the tangled filamentary structures that feature prominently outside the core. 

To verify that these filaments are indeed quantized vortices, we also plot slices of the phase and density profiles.
The vortical filaments pierce through the points at the centre of a $2\pi$ phase gradient and extremely low density regions, providing support to the interpretation of these structures as quantized vortices. This vortex tangle is made up from deformed vortex rings or, more accurately, topologically closed loops. 
In our simulations, vortex rings are formed which can expand, shrink and reconnect throughout the dynamical evolution~[\cite{2011JPhB...44k5101C,2021JCAP...01..011H}]. In a supplementary movie [\href{https://youtu.be/KXCFshb6-90}{SM Movie 2}], we present the full dynamical evolution for the halo displayed in Fig~\ref{fig:6}~(a). One can also readily observe there that the vortices in the inner halo move faster than the outer ones, which reconnect to each other rapidly.

In [\cite{2017MNRAS.471.4559M, 2021JCAP...01..011H}] it was mentioned that vortices appear outside the core. To quantify this statement further, in Fig~\ref{fig:6}~(a) we mark the region bounded by the scales of $r_c(t)$ and $r_t(t)$ and find that throughout the evolution the phase within $r<r_c$ is almost fully coherent, with the phase fluctuation increasing as we move outward to $r_t$ and beyond. From this observation, we indeed expect that  there will be no vortices passing within the core region since a vortex would be a source of phase incoherence and density fluctuations. To corroborate such result, supported by visual observation of our simulations, we introduce the local integral of the modulus square of the incompressible velocity field,
\be
\zeta(x,t)= \int_{V_c(x)}d\mb{r}|\mb{v}^i(\mb{r},t)|^2
\ee
where $V_c(x)$ stands for the volume of a spherical region of radius $r_c$, placed at different positions along the $x$ axis.
This quantity can qualitatively capture the incident rate of vortices crossing $V_c(x)$, a volume equal to that of the core but placed at various distances from the centre. Fig.~\ref{fig:6}~(b) shows $\zeta(x,t)$ for 4 different positions. For $r=0$, $\zeta(0,t)$ fluctuates around the values of $0.0344\pm0.0185$, while $\zeta(r_c,t)$ exhibits a similar pattern ($0.0498\pm0.0312$) (occasionally rising briefly to the order of $0.1$).
Moving towards the crossover region, the fluctuation rises significantly with values $0.5446\pm0.6960$, where the large variance indicates a considerable contribution to $\zeta$ when vortices pass through.
A similar pattern appears for $\zeta(x=10r_c,t)=0.4352\pm0.5811$ where we observe a peak value of comparable magnitude but a longer period of low values compared to $\zeta(x=r_t,t)$.
These features show that there are no vortices entering the soliton core region, or more precisely, the region $|\mb{r}|<r_c$ through time.

Around the crossover region there are several incidents of vortices crossing our sample volume, indicating a considerable population there. The vortex-crossing incident rate drops compared to the crossover region as we move towards the outer halo, in agreement with the incompressible energy density distribution in Fig.~\ref{fig:4}~(a).

\subsection{Superfluid turbulence in the outer halo}
%---------------------------------------
\begin{figure*}%[!ht]
    \centering
    \includegraphics[width=1\linewidth,keepaspectratio]{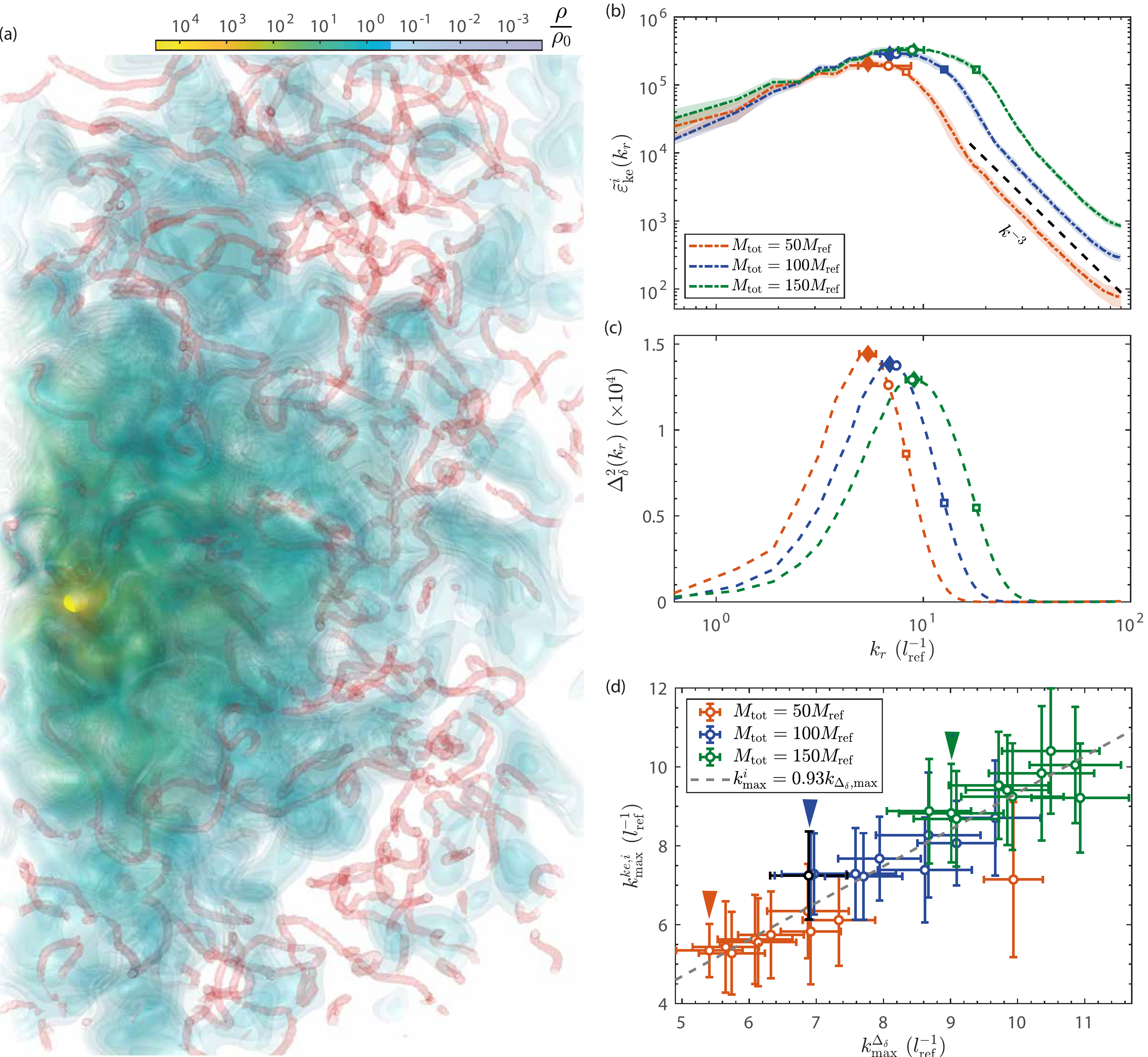}
    \caption{
    (a) 3D volume rendering for the density and the high vortical velocity field, displaying densities in the range  $[2.5\times10^{-5},0.5]\rho_c(t)$; this demonstrates that the typical inter-vortex distance coincides with the typical granule size with vortices winding in between %\gr{
    locally coherent%}
    density lumps.
    (b) The time-averaged incompressible kinetic energy spectra for halos with 3 different total masses. The spectrum peak indicates the typical inter-vortex distance. The black dashed line indicates the $k^{-3}$ scaling behaviour at high momenta which probes the vortex core structure - see text. 
    (c) The overdensity power spectrum for the same configurations as in (b).
    In (b) and (c) we show the momenta at which the 
    total density field $\rho(\mathbf{r},t)$ (hollow squares), overdensity $\delta\rho(\mathbf{r},t)$ (filled diamonds) and incompressible power spectra (hollow circles) respectively peak. 
    Note that the power spectrum for the total field $\rho(\mathbf{r},t)$ peaks at higher $k$ values (hollow squares) compared to the overdensity (filled diamonds). This is due to the somewhat smaller size of the core which contributes to the $\rho$ but not $\delta\rho$.
    The error bands in (b) illustrate the standard deviation from the time average, and the small horizontal error bars in both (b) and (c) show the standard deviation of the peak location over the averaged times. (d) The linear relation between $k_\mathrm{max}^{\Delta^2_\delta}$ and $k_\mathrm{max}^{\mathrm{ke},i}$ is shown with the black dashed line being the best fit from the scattered data giving $k_\mathrm{max}^{\mathrm{ke},i}=(0.9300\pm0.0274)k_\mathrm{max}^{\Delta^2_\delta}$. The outlier $M=50M_\mathrm{ref}$ simulation (bottom right) is the most bounded among our simulations with $E/M\approx-200$ and with a very dense core. The highlghted data points correspond to the simulations represented in (b) and (c) with our primary simulation represented by the black coloured data point. 
    }
    \label{fig:7}
\end{figure*}
%---------------------------------------

Turbulence in a quantum/superfluid, carried by quantized vortices, is a subject that has been extensively studied in non-self-gravitating atomic Bose-Einstein condensate systems~[\cite{1997PhFl....9.2644N,2014JSMTE..02..013T,Barenghi2014,Tsatsos_2016}], and in superfluid helium.
Turbulence is often regarded as a chaotic state but it can contain coherent structures in the tangle of vortex lines~[\cite{2014JSMTE..02..013T}].
In fact, during the equilibration of an ultracold quantum fluid from a highly non-equilibrium initial state, \cite{2002PhRvA..66a3603B} demonstrated the co-existence of dynamically-evolving vortices and coherent structures, with the long-range coherence of the system suppressed by the chaotic motions of the vortices (features also supported during the controlled cooling of a trapped ultracold superfluid~[\cite{2018CmPhy...1...24L}]). 
This view agrees qualitatively with the results depicted in Fig.~\ref{fig:3} and discussed in section \ref{sec:correlation_fns}, suggesting that the outer halo is indeed made up of locally coherent (density) structures.

Vorticity in FDM halos has been studied in [\cite{2017MNRAS.471.4559M}] and [\cite{2021JCAP...01..011H}] with the main conclusion of the earlier work [\cite{2011JPhB...44k5101C}], which did not include gravity, surviving the inclusion of the gravitational long-range force: vortices in such a system occur at places of destructive interference where the density can become zero, giving rise to the possibility of the phase winding by $2\pi$ around such regions. Interestingly, we note that, unlike cases of dynamically equilibrating superfluids in which efficient mode-mixing through the nonlinearity induced by the self-interaction ensures vortices can transfer energy into sound waves, the vortex structures in the non-self-interacting FDM case discussed here do not dissipate. We further comment on this below. 

One of the key quantities characterising quantum turbulence is the incompressible energy spectrum which coincides with the vortex energy spectrum,~[\cite{1997PhFl....9.2644N,1997PhRvL..78.3896N,2012arXiv1202.1863T,2012PhRvB..86j4501B,2012PhRvL.109t5304B,2014JSMTE..02..013T,2017JLTP..188..119T}]. 
Here we compute the incompressible energy spectrum in the form of~[\cite{1997PhFl....9.2644N}]
\be
\tilde{\epsilon}_\mathrm{ke}^{i}(k_r)\equiv\int d\mathbf{\Omega}\,k^2\tilde{\epsilon}_\mathrm{ke}^{i}(\mb{k})=\sum_{k_r\leq|\mb{k}|\leq k_r+\Delta k}\tilde{\epsilon}_\mathrm{ke}^{i}(\mb{k})
\ee
which is a cumulative measure of the amplitude of $\varepsilon_\mathrm{ke}^i(\mb{r})$'s (incompressible energy density) Fourier components for $k_r\leq|\mb{k}|<k_r+\Delta k$.

In Fig.~\ref{fig:7}~(b) we present the time-averaged $\tilde{\varepsilon}^i_\mathrm{ke}(k_r)$, together with the variance over time in shaded bands, for the three different total halo masses that we have simulated.
We can see that after $t=\tau_\mathrm{vir}$, the variance of $\tilde{\varepsilon}_\mathrm{ke}^i$ is marginal in time, suggesting a good convergence and that the system is in a steady state of turbulence, as first examined for the Schr\"{o}dinger equation in [\cite{2011JPhB...44k5101C}]. The location of the peak apparent in the incompressible kinetic energy spectrum, marked by hollow circles in Fig.~\ref{fig:7}~(b), signifies the characteristic intervortex length scale ~[\cite{2016PhRvA..94e3632S,2012PhRvB..86j4501B}]. As we discuss in the following subsection, this scale is also correlated to the typical granule size. In the high-$k$ region we also find a clear $k^{-3}$ behaviour, exhibiting a slight tilting around the grid resolution length scale. This last feature is an artefact of being too close to the smallest characteristic length scales of the system. 
The $\tilde{\varepsilon}^i_\mathrm{k}(k_r)\propto k^{-3}$ behaviour carries physical information and directly probes the vortex core structure~[\cite{1997PhFl....9.2644N, 2016PhRvA..94e3632S}], signifying that $\rho v^2\sim {\rm constant}$ on short scales\footnote{A power law  behaviour $r^{-s}$ leads to a $\sim k^{s-3}$ Fourier transform and, correspondingly, to a $\sim k^{2s-3}$ power spectrum.}. Given that quantized vortices must exhibit a $|\mb{v}^i|\propto r^{-1}$ velocity profile around the vortex cores (see Eqn.~(\ref{eq:vortical_velocity})) it follows that our measured spectrum signifies a $\rho\sim r^2$ profile for the local density around the core region as theoretically expected, see e.g.~[\cite{1997PhFl....9.2644N,2011JPhB...44k5101C}]. We note here that FDM turbulence was first numerically studied by \cite{2017MNRAS.471.4559M}\footnote{The turbulent features in FDM were also discussed in \cite{Woo:2008nn}.} through the spectrum of the full velocity for which they quoted a $k^{-1.1}$ behaviour. Although the authors of that work did not directly comment on this, this appears consistent with the expected velocity profile around the vortex cores.

A conspicuous feature of the incompressible kinetic energy spectrum is the absence of any segment exhibiting the Kolmogorov $k^{-5/3}$ scaling, usually associated with turbulent states. Furthermore, unlike most turbulence studies in laboratory Bose-Einstein condensates which show a decay of the turbulent state\footnote{The decay can be associated with an energy cascade~[\cite{2020AVSQS...2c5901M}] resulting in energy transfer from the system's incompressible component to the compressible one.} together with the decay of the number and line length of vortices via annihilations between vortex-antivortex pair and the emission of sound waves/phonons through Kelvin waves~[\cite{2003PhRvA..67a5601L,2005JPSJ...74.3248K,2016PhRvA..94e3632S}], FDM turbulence does not appear to decay in time as seen for example from the evolution of incompressible energy, see Fig.~\ref{fig:2}~(b). The emergence of a steady state that does not exhibit the Kolmogorov spectrum can be attributed to the absence of both a self interaction coupling all length scales via a nonlinear term, and of any source of dissipation, and was already recognised in [\cite{2011JPhB...44k5101C}] 
which studied the turbulence emerging in numerical solutions to the Schr\"{o}dinger equation. It appears that the addition of gravity only provides spatial differentiation of the density and does not alter the basic picture of this type of turbulence discussed in that work.       

%%%%%%%%%%%%%%%%%%%%%%%%%%%%%%%%%%%%%%%%%%%%%%%%%%%%%%%%%%%%%%
\subsection{Granule spectrum}
Lastly, we look at the interplay of turbulence and the structural opposite of vortices, those regions of constructive interference that, as we discussed above, exhibit local density coherence and are commonly referred to as granules in the FDM literature. 
To characterize the density fluctuations/granules, we reformulate the density as~[\cite{2018MNRAS.478.2686C,2018PhRvD..97j3523L,2021ApJ...916...27D}]
\be
\rho(\mathbf{r},t)=\bar{\rho}(r)\left[1+\delta\rho(\mathbf{r},t)\right]
\ee
where now $\bar{\rho}(r)$ is the spherically symmetric core-halo density profile, obtained by the azimuthal average of the time averaged density, $\langle\rho(\mathbf{r})\rangle_t$.\footnote{$\bar{\rho}$ in this section should not be confused with the use of the same symbol in earlier sections which denotes a different average.%\gr{GARY: IS THIS ACCURATE?}
}
Here the overdensity is defined by
\be
\delta\rho(\mb{r},t)=\frac{\rho(\mb{r},t)-\bar{\rho}(r)}{\bar{\rho}(r)}
\label{eq:density_fluc}
\ee
where $\rho(\mb{r},t)$ is re-centred by setting $\rho(\mb{r}=0)$ as the peak density and gives the density fluctuations on top of the spherical $\bar{\rho}(r)$.
The azimuthal avaraging aims to reduce artefacts on the identification of the granule size from very slow moving local structures that could remain in $\langle\rho(\mathbf{r})\rangle_t$ after temporal averaging as well as from the random motions of the soliton core.

In order to to probe the characteristc granule length scales~[\cite{2018MNRAS.478.2686C,2021ApJ...916...27D,2018PhRvD..97j3523L}] and similar to the discussion on the core, we consider the power spectrum of the overdensity,
\ba{rl}
\Delta^2_\delta(k_r,t)=&\displaystyle\frac{k_r^3}{(2\pi^2)4\pi k_r^2}\int d\mb{\Omega}_kP_\delta(\mb{k},t)
\\\\
\approx&\displaystyle\frac{k_r^3}{2\pi^2 \mathcal{N}_{k_r}}\sum_{k_r\leq|\mb{k}|<k_r+\Delta k}|\mb{k}|P_\delta(\mb{k},t)
\ea
where $P_\delta(\mb{k},t)=|\tilde{\eta}_\delta(\mb{k},t)|^2$ and $\tilde{\eta}_\delta(\mb{k},t)$ is the Fourier transformation of the overdensity $\delta(\mb{r},t)$.
In Fig.~\ref{fig:7}~(c) we show the time averaged power spectrum of the overdensity for the same set of data as in (b). The peak locations of the overdensity power spectrum are marked as filled diamonds in (c) and are also marked in (b). In Fig.~\ref{fig:7}~(d) we plot the relation of $k_\mathrm{max}^{\Delta^2_\delta}$ and $k_\mathrm{max}^{\mathrm{ke},i}$ for our 30 simualtions exhibiting a clear linear relation between these two quantities. The best fit for the scattered data is  
\be
k_\mathrm{max}^{\mathrm{ke},i}=(0.9300\pm0.0274)k_\mathrm{max}^{\Delta^2_\delta}
\ee 
showing the typical granule size to be essentially equal or marginally smaller than the typical intervortex distance. 
This is consistent with the picture of granules being regions of constructive interference with vortices surrounding them, residing in low density regions of destructive interference. Fig.~\ref{fig:7} (a) shows a 3D volume rendering for the density and the regions identified as vortices via their high velocity values, visually illustrating that idea that vortices surround the granules in a FDM halo, as first pointed out in [\cite{2021JCAP...01..011H}].

\section{Discussion and Conclusions}
\label{sec:conclusion}
In this work, we have carried out 30 simulations of virialized FDM halos, generated by the merging of randomly located coherent solitons. The final products of our merger numerical experiments contain the characteristic FDM solitonic cores at the centre and exhibit the striking FDM granular structure outside them, which, when averaged over time and angles, follows an $r^{-3}$ trend consistent with the NFW profile. We then examined these halos through the prism of the field's coherence, employing measures widely used in the study of laboratory Bose-Einstein condensates: the lowest eigenmodes of the single-particle density matrix $\rho(\mathbf{r},\mathbf{r'})$, the correlation functions $g_1$ and $g_2$ as well as the reduced phase space density. 

All coherence measures are compatible with a bimodal spatial configuration that involves a central, fully coherent Bose-Einstein condensate that aligns perfectly with the soliton, embedded in a turbulent halo that is on average incoherent but appears to exhibit local (spatially and temporally) coherent quasi-condensate patches, with an identifiable crossover region between the solitonic core and (outer) halo. The configuration is in many respects analogous to finite temperature laboratory BECs confined in a (harmonic) potential where the condensate sits at the bottom of the potential and is surrounded by a cloud of incoherent (thermal) particles. Here, the role of the trap is played by the gravitational potential of the whole halo. It is also worth stressing that, although the particle number is very large and fully justifies a classical field approach, the \emph{reduced} phase space density beyond the crossover region drops below unity, indicating the possibility that a description via distinguishable, individual particles may capture some features of the halo dynamics outside the core.    

The identification of the central soliton with a perfect BEC state was obtained via the numerical determination of the Penrose-Onsager mode of the field configuration (mode with largest eigenvalue of the single-particle density matrix) which coincides with the empirically derived soliton profile. The outer halo was deemed incoherent on average. However, a more detailed inspection of the $g_2$ correlation function, when averaged over differing timescales, shows that some density coherence remains over short temporal intervals and localised in space. These features may be tentatively identified with the "granular texture" often employed to describe the appearance of the outer parts of FDM halos. Precise characterization of these features though would require more extensive numerics and should be left for future work. 

We found that the "core-crossover region-outer halo" spatial structure implied by the measures of field coherence doesn't only reflect static or average properties but also describes the dynamical oscillations of the core.\footnote{We have also confirmed the random motion of the central core reported in ~[\cite{2021PhRvD.103b3508L,2020PhRvL.124t1301S,2021ApJ...916...27D}].} Interestingly, these oscillations were found to be anti-correlated to the peak location of the power spectrum of the whole halo. This would  suggest that either the core leaves its imprint on the field fluctuation power spectrum, at least for the halos in our numerical experiments, or that the typical size of the density fluctuations, the granules, varies with time. Clarification of this would be another topic for future work but preliminary results indicate that the granule size appears to remain fixed with time, suggesting the core is likely driving the power spectrum oscillation. 

The phase incoherence of the halo outside the core is expressed via the formation of a tangle of quantum vortices, surrounding the phase and density semi-coherent granules and exhibiting sustained quantum turbulence with persistent, non-decaying, incompressible kinetic energy. We identified filamentary structures as regions of high velocity and confirmed that they are indeed quantum vortices by cross-checking with the local phase profiles around them. These vortices are confined outside the radius of the solitonic core. We then examined the power spectrum of the incompressible kinetic energy which we found to exhibit a characteristic peak that marks the intervortex distance. Its location was found to essentially coincide with the peak of the granule power spectrum, which in turn corresponds to a characteristic granule size. Thus, we have explicitly found that vortices do indeed form in between the granules. Furthermore, vortices are not ideal line-like defects but have a specific width over which density and velocity develop characteristic profiles. We found the incompressible kinetic energy spectrum to exhibit a $k^{-3}$ scaling in the ultraviolet momentum region which, given that the quantum vortex velocity profile has to be $|\mb{v}^i|\propto r^{-1}$, implies a density $\rho \sim r^2$ around the vortex cores. It might be worth noting that the vortex energy spectrum of the halos in our simulations do not exhibit the well known Kolmogorov $k^{-5/3}$ scaling, which is associated with the bundles of vortex lines on length scales larger than the vortex size~[\cite{2012PhRvL.109t5304B}].

Before closing let us make a few comments regarding the physical scales for which our simulations can be applied. Considering a fiducial boson mass of $m=2.5\times10^{-22}$~eV/$c^2$ and a fiducial mass density of $10^3M_\odot\textrm{kpc}^{-3}$, comparable to the mean cosmic value (and noting that the estimated average dark matter density in a halo hosting a milky way-type galaxy is approximately
$\rho_\mathrm{Halo}\approx 10^7 M_\odot\textrm{kpc}^{-3}$) our reference units can be written as 
\be
E_\mathrm{ref}\approx 5.38\times10^{-27}\left(\frac{\rho_\mathrm{ref}}{10^3M_\odot\textrm{kpc}^{-3}}\right)^{1/2}\mathrm{ eV}\;,
\ee
\be
l_\mathrm{ref}\approx 10.81\left(\frac{2.5\times10^{-22}\mathrm{ eV}}{mc^2}\right)^{1/2}\left(\frac{10^3M_\odot\textrm{kpc}^{-3}}{\rho_\mathrm{ref}}\right)^{1/4}\mathrm{ kpc}\;,
\ee
and
\be
\tau_\mathrm{ref}\approx 14.9\left(\frac{10^3M_\odot\textrm{kpc}^{-3}}{\rho_\mathrm{ref}}\right)^{1/2}\mathrm{Gyr}\;,
\ee
with 
\ba{rl}
M_\mathrm{ref} =& \rho_\mathrm{ref} l_\mathrm{ref}^3 \\\\ \approx&\displaystyle 1.26\times 10^6M_\odot\left(\frac{2.5\times10^{-22}\mathrm{ eV}}{mc^2}\right)^{3/2}\left(\frac{\rho_\mathrm{ref}}{10^3M_\odot\textrm{kpc}^{-3}}\right)^{1/4}\;.
\ea
The reference velocity in our simulations,  $v_\mathrm{ref}=l_\mathrm{ref}/\tau_\mathrm{ref}$, is $0.71$ km/s for $\rho_\mathrm{ref}=10^3 M_\odot\mathrm{kpc}^{-3}$, or  e.g.~$3.99$ km/s for $\rho_\mathrm{ref}=10^6 M_\odot\mathrm{kpc}^{-3}$, and the velocities of the FDM particles in the outer parts of the halo are measured to be $5$ to $10$ times that reference velocity in the total mass range of $50 M_\mathrm{ref}$ to $150M_\mathrm{ref}$. We thus see that the halos in our merger simulations are "colder" than what would be expected for typical dark matter halos hosting a Milky Way-type galaxy and therefore expect the features discussed in this paper to be present at relatively smaller scales within the halo in more realistic conditions. The reference phase space density is 
\ba{rl}
{\cal D}_\mathrm{ref}=&\displaystyle\frac{h^3\rho_\mathrm{ref}}{m^4v_\mathrm{ref}^3}
\\\\
\approx&\displaystyle1.25\times10^{90}\left(\frac{10^{-22}\textrm{eV}/c^2}{m}\right)^4\left(\frac{\rho_\mathrm{ref}}{10^3M_\odot\mathrm{kpc}^{-3}}\right)\left(\frac{250\mathrm{km/s}}{v_\mathrm{ref}}\right)^3
\ea
where, the value of 250 km/s was chosen as the typical velocity of a particle in a Milky-way-like environment. Because the boson mass relevant to FDM is so small, the resulting  number of particles is enormous for our parameters, and the condensation criterion (\ref{cond_crit-1}) is amply satisfied throughout our halos. This of course amply justifies the use of a wave equation.

Although we performed 30 randomly generated simulations, the configurations we examined gave very similar sizes of core radii while the initial energy configurations involved only about a factor of 5 difference. Furthermore, as discussed above all our configurations are less energetic than typical dark matter halos of comparable masses. In addition, the periodic boundary condition may play a role in the overall halo while the propagating waves could reenter the system. The natural next step would therefore be to apply the coherence and quantum turbulence analysis of this work to more realistic halos obtained from a bigger simulation with cosmological initial conditions. Moreover, it would be interesting to include self-coupling and also investigate possible observational signatures of these features which would be specific to FDM halos. We will focus on these topics in future works. 

\section*{Acknowledgements}

I.K.L. acknowledges funding from European Union’s Horizon 2020 research and innovation programme under the Marie Sklodowska-Curie grant agreement No. 897324 (upgradeFDM), while G.R. and N.P.P. acknowledge funding from the Leverhulme Trust (Grant no. RPG-2021-010).
I.K.L. gratefully thanks Professor Hsi-Yu Schive and Professor Tzihong Chiueh for their hospitality and enlightening discussions during his extended visit to their group which contributed towards the precise identification of some of the questions addressed here and which allowed us to crystallize our thinking on them. We also acknowledge indirect contributions by Milos Indjin, and discussions with Carlo Barenghi, Piotr Deuar, Luca Galantucci, Cora Uhlemann and Alex Gough.
The simulations were conducted on the HPC Cluster Rocket of Newcastle University. %Data supporting this publication can be openly accessed under an `Open Data Commons Open Database License' on the  site~[\href{https://data.ncl.ac.uk:12604728}{data.ncl.ac.uk:12604728}].

\section*{Data Availability}
Data supporting this publication can be openly accessed under an `Open Data Commons Open Database License' on the  site~[\href{https://doi.org/10.25405/data.ncl.22004978}{https://doi.org/10.25405/data.ncl.22004978}].

%\bibliographystyle{mnras}
%\bibliography{ref_GR,ref_final} % if your bibtex file is called example.bib

\appendix

\section{Scaling Transformations of Schr\"{o}dinger--Poisson Equations}
\label{appendix:scaling}
The SPE are invariant under the Lifshtz-type transformation~\cite{1994PhRvD..50.3650S,2004PhRvD..69l4033G},
\be
\left\{
t,x,\Phi,\psi,\rho
\right\}\rightarrow\left\{
\Lambda^{-2} \hat{t},\Lambda^{-1}\hat{x},\Lambda^{2}\hat{\Phi},\Lambda^{2}\hat{\Psi},\Lambda^{4}\hat{\rho}
\right\}\,,
\label{eq:scaling_1a}
\ee
With the total mass, energy and angular momentum of the system transforming as
\be
\left\{
M,E,L
\right\}\rightarrow\left\{
\Lambda\hat{M},\Lambda^{3}\hat{E},\Lambda\hat{L}
\right\}
\label{eq:scaling_1b}\,.
\ee
Furthermore, the SPE can also be scaled by the boson mass, $m\rightarrow\alpha m'$ as~[\cite{2017MNRAS.471.4559M,2021PhRvD.103b3508L}],
\be
\left\{
t,x,\Phi,\psi,\rho
\right\}\rightarrow\left\{
\alpha\hat{t},\hat{x},\alpha^{-2}\hat{\Phi},\alpha^{-1}\hat{\Psi},\alpha^{-2}\hat{\rho}
\right\}
\label{eq:scaling_2a}
\ee
and, accordingly,
\be
\left\{
M,E,L
\right\}\rightarrow\left\{
\alpha^{-2}\hat{M},\alpha^{-3}\hat{E},\alpha^{-2}\hat{L}
\right\}
\label{eq:scaling_2b}.
\ee
and the combination of the above transformations leads to
\begin{equation}
\left\{
t,x,\Phi,\psi,\rho
\right\}\rightarrow\left\{
\Lambda^{-2}\alpha\hat{t},\Lambda^{-1}\hat{x},\Lambda^2\alpha^{-2}\hat{\Phi},\Lambda^2\alpha^{-1}\hat{\Psi},\Lambda^4\alpha^{-2}\hat{\rho}
\right\}
\label{eq:scaling_3a}
\end{equation}
\begin{equation}
\left\{
M,E,L
\right\}\rightarrow\left\{
\Lambda\alpha^{-2}\hat{M},\Lambda^3\alpha^{-3}\hat{E},\Lambda\alpha^{-2}\hat{L}
\right\}
\label{eq:scaling_3b}.
\end{equation}
These scaling properties would allow one to use an individual simulation for describing configurations with different physical parameters by appropriately scaling the reference units. 

\section{Numerical simulation details}
\label{appendix:numeric}
\subsection{Numerical methods}
\label{appendix:B1}

Defining the dimensionless variables (primes omitted in the main text)
$t'=t/\tau_\mathrm{ref}$, $\mb{r}'=\mb{r}/L_\mathrm{ref}$, 
$\Psi'=\Psi/\sqrt{\rho_0}=\Psi/\sqrt{\varrho^\prime\rho_\mathrm{ref}}$, 
$\varrho^\prime=\rho_0/\rho_\mathrm{ref}$,
where
$\tau_\mathrm{ref}=\displaystyle(1/\sqrt{G\rho_\mathrm{ref}})$,
and 
$l_\mathrm{ref}=\displaystyle\left(\hbar^2/m^2G\rho_\mathrm{ref}\right)^{1/4}\,$,
the dimensionless form of the SPE reads,
\ba{rl}
\displaystyle i\frac{\partial}{\partial t'}\Psi'(\mathbf{r}',t')=&\displaystyle\left[-\frac{\nabla'^2}{2}+\Phi'(\mathbf{r}',t')\right]\Psi'(\mathbf{r}',t')
\\\\
\nabla'^2\Phi'(\mathbf{r}',t')=&4\pi\varrho^\prime\left(|\Psi'(\mathbf{r}',t')|^2-1\right)\;.
\ea
The normalization of $\Psi'$ is considered to be the volume of the simulation box and the total mass,
\be
V'=\int d\mb{r}'|\Psi'|^2\quad\textrm{and}\quad M'_\mathrm{tot}=V'\varrho^\prime\;.
\ee

The Poisson equation is solved using the pseudo-Fourier spectrum method with the implementation of discrete Fourier transformation,
\be
\Phi'(\mb{r}',t')=-\mathcal{F}^{-1}\left[-\frac{1}{k^2}\left\{\tilde{\rho'}(\mathbf{k}')-\varrho^\prime\delta(\mb{k}')\right\}\right]
\ee
where $\mathcal{F}[\cdots]$ and $\mathcal{F}^{-1}[\cdots]$ are the (discrete) Fourier and inverse Fourier transformations and are given by
\be
\tilde{f}(\mb{k}')=\mathcal{F}[f(\mb{r}')](\mb{k})=\sum_{\mb{r}'}e^{-i\mb{k}'\cdot\mb{r}'}f(\mb{r}')\Delta x'\Delta y'\Delta z'
\ee
and
\be
f(\mb{r}')=\mathcal{F}^{-1}[\tilde{f}(\mb{k}')]=\sum_{\mb{k}'}\frac{e^{i\mb{k}'\cdot\mb{r}'}}{V'}\tilde{f}(\mb{k}')
\ee
respectively, with $\tilde{\rho'}(\mathbf{k}')=\mathcal{F}\left[|\Psi(\mb{r}',t')|^2\right](\mathbf{k}')$.
The divergence of $k^{-2}$ at $\mb{k}=0$ is eliminated by the subtraction of the averaged density because $\mathcal{F}^{-1}[\mathcal{F}[\rho'(\mb{r}')](\mb{k}=0)]=(1/V')\sum_{ijk} \rho'(\mb{r}'_{ijk})=\bar{\rho}'$.

The dimensionless SPE is propagated by the second order time-splitting method, also known as kick-drift-kick (KDK) method, which has been widely implemented in previous studies~[\cite{2014NatPh..10..496S, 2017MNRAS.471.4559M,2021ApJ...916...27D}].
This method decomposes a Hamiltonian of a Schr\"odinger-like equation into the kinetic and potential energy terms to approximate the propagator as
\be
\hat{P}(\Delta t')=e^{i\hat{H}'\Delta t'}\approx e^{-i\hat{K}'\Delta t'/2}e^{-i\hat{V}'\Delta t'}e^{-i\hat{K}'\Delta t'/2},
\ee
where $\hat{K}=-\nabla'^2/2$ and $\hat{V}'=\Phi'$;
such approximation is accurate to $\mathcal{O}(\Delta t'^3)$. The wavefunction is evolved from $t'$vto $(t'+\Delta t')$ via the propagator, $\Psi'(\mb{r}',t'+\Delta t')=\hat{P}(\Delta t')\Psi'(\mb{r}',t')$.
Recent works have also adapted different higher-order methods including higher-order time-splitting method~[\cite{2021PhRvD.104h3022C,2020PhRvD.102h3518S}] and 4th-order Runge-Kutta method~[\cite{2016PhRvD..94d3513S}].
One of the great advantages of the method used here is that normalization is perfectly preserved, namely, $d V'/dt'=0$, with this series of unitary operations.
To ensure numerical stability, the time step is constrained by he Courant-Friedrichs-Lewy~(CFL)-like condition~[\cite{Ajaib,2017MNRAS.471.4559M}],
\be
\Delta t' \leq \Delta t_\mathrm{max}' \;, \hspace{0.2cm} {\rm where} \hspace{0.2cm} 
\Delta t_\mathrm{max}' = \min\left[\frac{(\Delta x')^2}{6},\frac{1}{|\Phi|'_\mathrm{max}}\right],
\ee
requiring the change of phase after each kick and drift unitary transformation to be less than $2\pi$.
The grid spacing is set by the number of grid points $N_x=N_y=N_z$, which, along with the length of the computational box $L_x'=L_y'=L_z'=L^\prime$, give $\Delta x'=L_x'/N_x$.
The CFL-like condition shows that the timestep scales as $(\Delta x^\prime)^2$ rather than the $\Delta x^\prime$ for gravity and the Eulerian fluid solver, 
which adds computational costs to simulations.
We chose a value of $\Delta t'=10\times^{-2}\Delta t^\prime_\mathrm{max}$ to ensure that the total energy loss is less than $1.6\%$ throughout the entire probed dynamical evolution; in our simulations, we also noticed that the energy loss is correlated to the total mass set by $\varrho^\prime$ in our SPE solver.
In general, to achieve same level of accuracy, an even smaller fraction of $\Delta t^\prime_\mathrm{max}$ is needed when $M_\mathrm{tot}$ is heavier, or $\varrho^\prime$ is larger.

\subsection{Time-average period convergence}
\label{appendix:data}
%---------------------------------------
\begin{figure}%[!ht]
    \centering
    \includegraphics[width=1\linewidth,keepaspectratio]{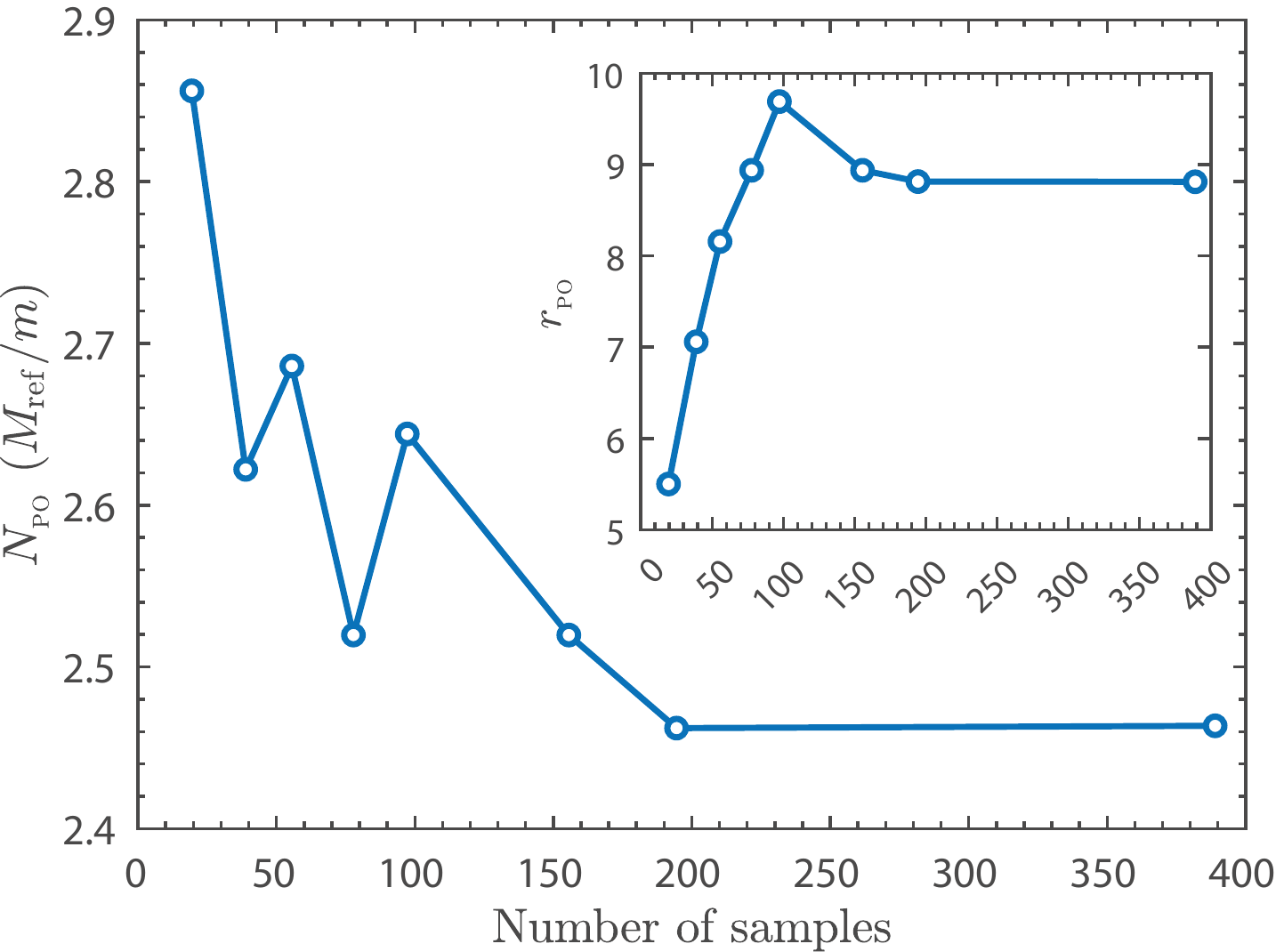}
    \caption{
    The particle numbner of the Penrose-Onsager mode for the density matrix averaged over a different number of samples (temporal snapshots) during the post-virialized evolution, of the time span %$4.8625\tau_\mathrm{ref}$ 
    $t\in[4.5,9.3625]\tau_\mathrm{ref}$ for the primary data presented in this work. The inset shows the ratio of the first (PO) to the second largest eigenvalues of the density matrix, $r_{\mathrm{PO}}=N_0/N_1$, confirming the numerical averaging convergence used in this work, and the establishment of a single macroscopically-occupied eigenvalue depicting the condensate mode.
    }
    \label{fig:A1}
\end{figure}
%---------------------------------------

Results shown in the main text as a function of the radial coordinate, $r$, typically involve both temporal and angular averaging. Let us discuss these here. The idea of using temporal averaging to extract averaged quantities replaces the need for ensemble averaging which, while preferential in principle, would be numerically prohibitive, and becomes strictly valid under the assumption of ergodicity for the system. After virialization, we have clearly identified features which remain effectively (on average) unchanged in time, like the system density and solitonic core properties, and other features which still vary on a non-negligible timescale due to the dynamical/turbulent evolution of vortices. Aside from the total averaging timescale, one needs to ensure that enough samples are being averaged over to obtain sufficiently smooth features.

To proceed with averaging in any given individual numerical simulation, we first probe the system to ensure that its energy has reached a steady-state: this is evidently well satisfied after $\tau_{\rm vir}$ (see Fig.~\ref{fig:2}(b) in main text). Beyond such time, we need to carefully select our averaging procedure.

In order to extract the radial profile, and compare both to the solitonic core typically used for the inner core region and the NFW profile at sufficiently larger distances, we average over the entire time duration -- which corresponds to $\approx 4.9 \tau_{\rm ref}$ and $\approx 390$ snapshots during the system's evolution. As evident from [\href{https://youtu.be/KXCFshb6-90}{SM Movie 2}] such timescale is long enough to wash out any internal system dynamics, to obtain a smooth density profile. In order to improve on averaging, 
we also integrate over the azimuthal direction, to look at radial shells from the centre of the solitonic core. This has allowed us, for example, to construct the radial density profiles (which we have fit with the bimodal profile of Eq.~(\ref{eq:core-halo_fit})) of Figs.~\ref{fig:2}(c) and~\ref{fig:3}(a), the PO mode and next few eigenmodes, shown in Fig.~\ref{fig:3}(c), and the radial evolution of the different (decomposed) energy contributions.
As an illustration of such averaging process, we highlight here [Fig.~\ref{fig:A1}] the convergence of the two largest eigenvalues of the density matrix with the number of temporally-successive samples (field snapshots) being averaged over. We can clearly see that an average of 200 samples is enough to accurately probe both the PO mode [Fig.~\ref{fig:A1}], as well as the mode with the next highest eigenvalue -- the latter is evidenced here by monitoring the ratio $r_{\rm PO}=N_0/N_1$ of the largest to the second largest eigenvalue [see inset to such figure]. Nonetheless, to guarantee sufficient convergence, our results are typically based on averaging over a much larger number of modes (with at least 310 snapshots for any individual simulation), and nearly double the time-span.

Our analysis has indicated that, while the core properties are largely time-independent, the outer halo ($r>r_t$) exhibits rather intricate dynamics, through the chaotic motion of quantized vortices. In order to probe time-averaged density fluctuations ($g_2(r)$) in such regions, one would need to average over timescales much longer than those of typical vortex evolution.
Our simulations demonstrate that typical vortex dynamical timescales are in fact dependent on the radial coordinate [\href{https://youtu.be/KXCFshb6-90}{SM Movie 2}], with motion being slower in the outer parts of the halo, qualitatively consistent with the finding in~[\cite{2018PhRvD..97j3523L}]. As such, large density fluctuations associated with vortices can persist on rather long timescales, and this could be the reason why we find $g_2 \gtrsim 2$ even after full temporal (and angular) averaging, as seen in Fig.~\ref{fig:3}(b) and~\ref{fig:3}(d)(ii). As noted in the main text, this feature makes the identification of the quasi-condensate via Eq.~(\ref{eq:QC}) rather numerically problematic for $r_t$. Nonetheless, our detailed statistical analysis over 30 different initial conditions simulations featuring 3 different total mass configurations (see Appendix~\ref{appendix:all_data}), shows that the average value of $g_2 \approx 2$ at $r \approx 10r_c$ across all such 30 simulations (although the large error bar  does not allow us to 
completely rule out non-Gaussian fluctuations manifesting themselves in our model).

Although numerically it is hard to obtain a more accurate value of $g_2$  ($\approx 2$)in regions where density fluctuations dominate due to the chaotic propagation of vortices through all outer halo regions, the absence of any values of $g_2 \ll 2$, signifying suppressed density fluctuations, does mask the actual shorter-time evolution of individual snapshots. On individual times/snapshots (rather than time-averaged findings), our simulations consistently indicate the existence of regions of slowly-varying densities -- both above and below the mean density at a given radius -- separated by vortices. Such regions in fact only start exhibiting significantly distinct features (associated with motion of vortices) beyond a timescale of $\sim O(0.1 \tau_{\rm ref}$). 
As such, we have also probed $g_2$ features over a timescale of $0.5 \tau_{\rm ref}$ -- corresponding to 40 snapshots of the field configuration -- which is enough both to get reasonable averaging, and to probe the configuration of the field over relatively short evolution times.
While such results, shown in the left column of Fig.~\ref{fig:3}(d), still have problems with $g_2 \gtrsim 2$, importantly in this case we see very clear signatures of local suppression of density fluctuations over a non-negligible temporal duration: these are shown in Fig.~\ref{fig:3}(d)(iii), revealing a large number of randomly distributed regions of sizes broadly comparable to the solitonic core radius with $g_2 < 1.2$ (but still clearly $>1$, distinguishing such features from those of the solitonic core), a clear signature of locally reduced density fluctuations over a relatively extended time domain. To understand this better, we have also probed the spatial correlation around the centre of some of these overdense/underdense regions (compared to the average density at such radial locations), confirming that such regions of noticeably suppressed density fluctuations also exhibit suppressed phase fluctuations. In other words, the outer halo does consist of dynamically-evolving regions of (somewhat) reduced density and phase fluctuations, but the turbulent nature of the vortex dynamics obscures such features unless one restricts the analysis to timescales which are shorter than, or comparable to, those of the evolution of quantum vortices within the tangle.

\subsection{Independence of Findings on Grid Resolution}

%---------------------------------------
\begin{figure}%[!ht]
    \centering
    \includegraphics[width=1\linewidth,keepaspectratio]{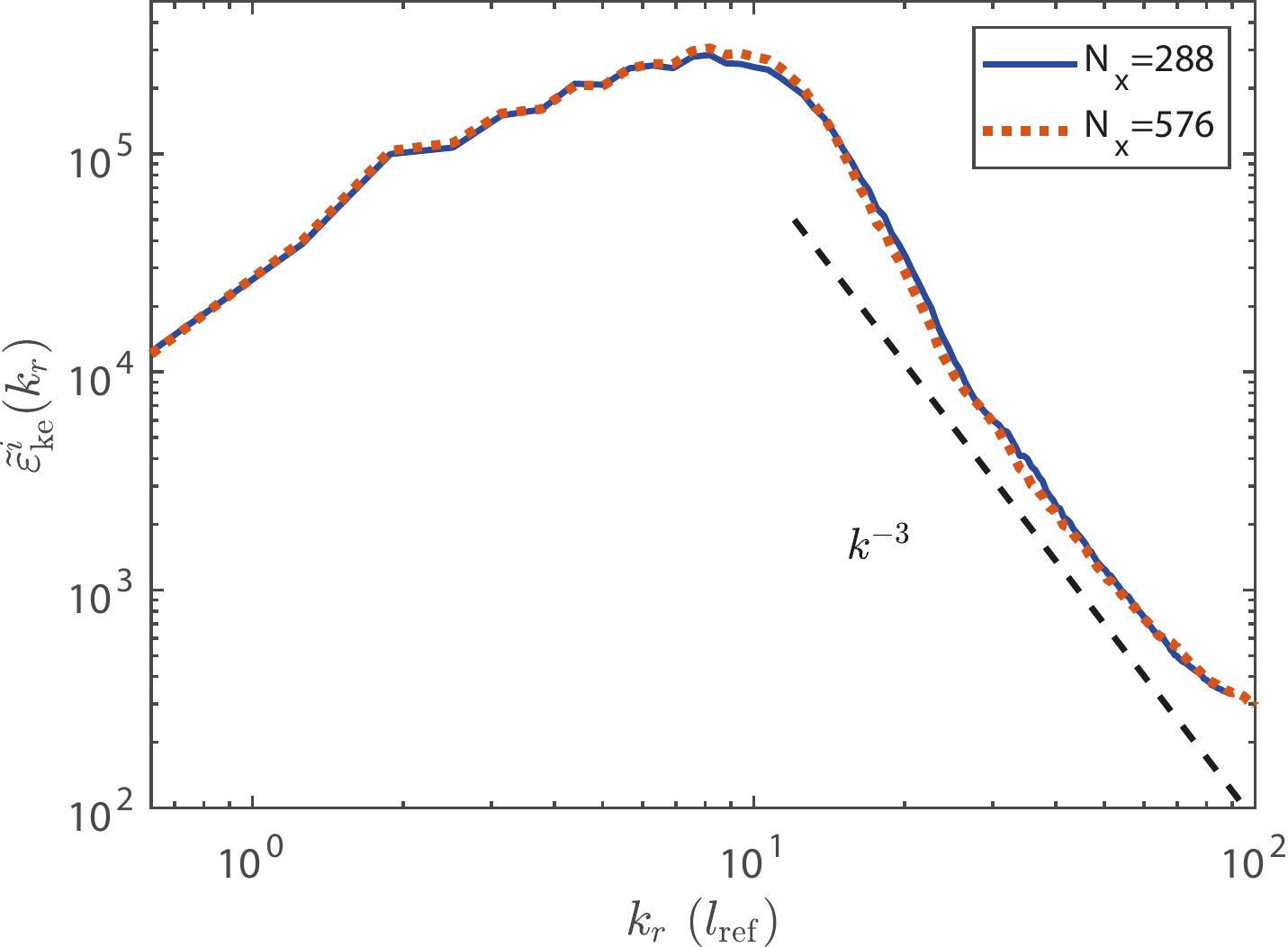}
    \caption{
    Comparison of the incompressible kinetic energy spectrum for the primary dataset presented in this work for our standard grid number $N_x=288$ ($\Delta x\approx0.035l_\mathrm{ref}$), vs.~corresponding result for the double grid $N_x=576$ ($\Delta x\approx0.017l_\mathrm{ref}$) for results taken at $t=8\tau_\mathrm{ref}$. 
    }
    \label{fig:A2}
\end{figure}
%---------------------------------------
To be sure of the convergence of our results and their independence to the numerical grid resolution, we have doubled our spatial resolution (i.e.~halved our numerical spatial disretization) and verified that our findings remain unchanged. We demonstrate this here specifically for the case of the incompressible kinetic energy spectrum.
To compare the results as closely as possible, we interpolate the wavefunction of our primary simulation at $t=7.625\tau_\mathrm{ref}$ from $N_x=288$ to $N_x=576$ for the same box size, $L_x=10l_\mathrm{ref}$, and then propagate it for a short amount of time to eliminate any artefacts introduced through our implemented grid interpolation.
The result of the incompressible energy spectrum at $t=8\tau_\mathrm{ref}$ is shown in Fig.~\ref{fig:A2}: these clearly demonstrate
a similar momentum window exhibiting the $k^{-3}$ scaling for the vortex structure. Scaling behaviour is expected to appear in the region $2\pi/L< k<2\pi/\xi$ where $L$ is the inter-vortex distance length scale, characterized by the wavenumber for which the spectrum peaks, and $\xi$ is the typical vortex length scale of the system~[\cite{1997PhFl....9.2644N,2016PhRvA..94e3632S}].

\section{Statistical Analysis of Measured Coherence Properties over Different Initial Configurations}

\label{appendix:all_data}

\begin{figure}%[!ht]
    \centering
    \includegraphics[width=1\linewidth,keepaspectratio]{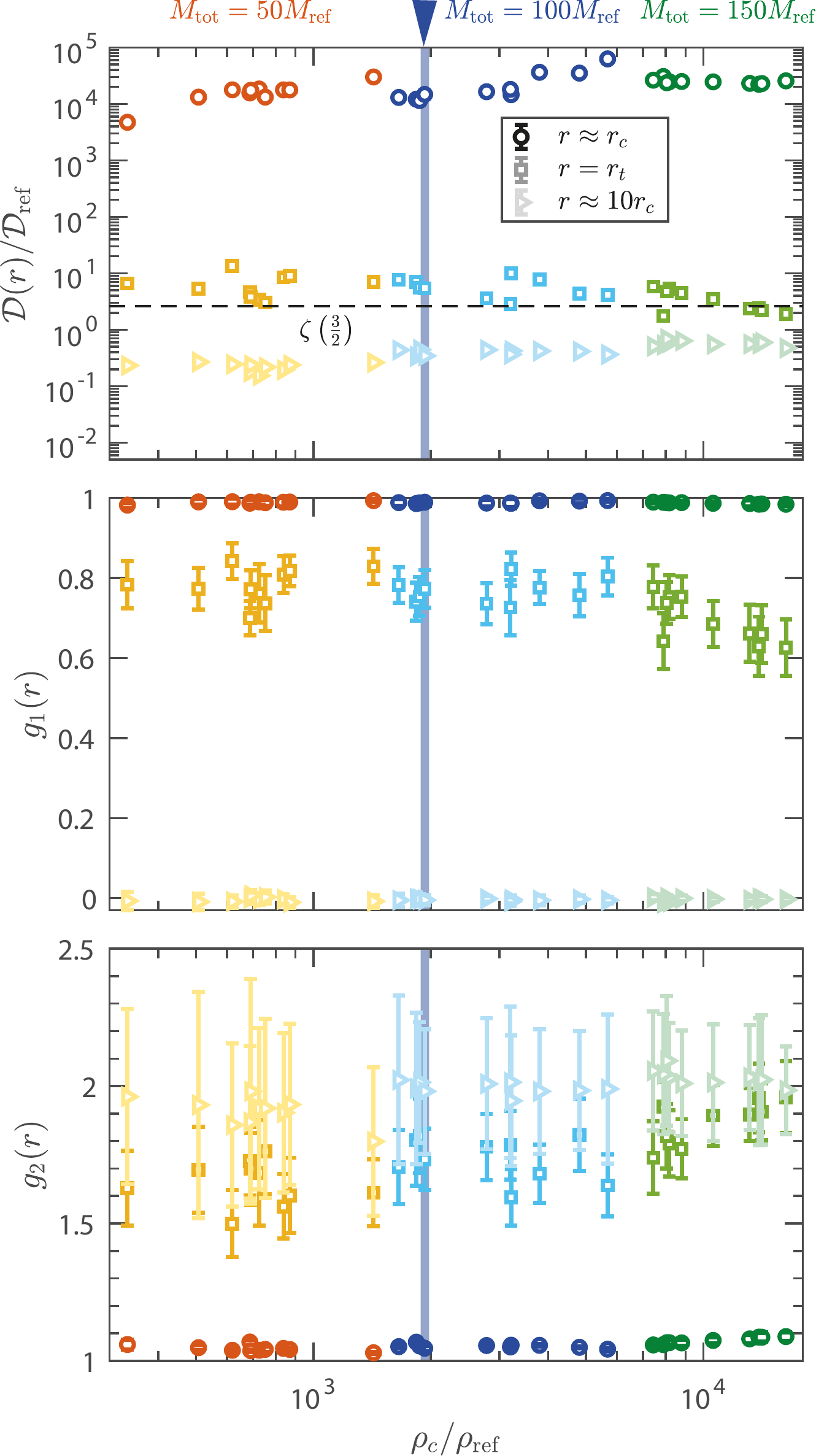}
    \caption{
    Values of distinct measures related to system coherence evaluated at the solitonic radius $r_c$ (circles), the (outer) crossover
    radius $r_t$ (squares), and a further point in the outer halo  with $r=10 r_c$ (triangles)
    for different total masses as a function of (scaled) core density $\rho_c/\rho_{\rm ref}$. 
Shown are (a)    the scaled phase space density ${\cal D}(r)/{\cal D}_{\rm ref}$; (b), $g_1(r)$ and, (c) $g_2(r)$, with such measures clearly highlighting the (gradual) onset of coherence for $r < r_t$, with the system acquiring practically total coherence (with a dimensionless phase-space density exceeding $~O(1)$) for $r \le r_c$. The black dashed line in (a) indicates the critical phase space density value for the homogeneous ideal Bose gas.
The blue band marked by the blue triangle points the primary data presented in the main text.
    }
    \label{fig:A3}
\end{figure}

In this work, we have analyzed density and coherence properties of a particular soliton merger initial condition, and used this as a basis for a number of conclusions about the physics of a single solitonic core and its surrounding halo. Although the main text has focused on a single simulation, the obtained features are generic, and independent -- within our range of probed total mass and energy (see Fig.~\ref{fig:2}(d))-- of initial conditions. To confirm such independence, Fig.~\ref{fig:A3} shows the values of our dimensionless phase-space density, ${\cal D}(r)/{\cal D}_{\rm ref}$, and coherence measures, $g_1(r)$ and $g_2(r)$ at 3 different values of $r$ corresponding to the soliton core radius, $r_c$, the outer crossover region radius, $r_t$, and at a much larger radial distance of $10 r_t$ in the outer halo. These are shown for different total mass configurations $M_{\rm tot}/M_{\rm ref}=$ 50 (orange, leftmost set of points) , 100 (blue, middle set of points) and 150 (green, rightmost set of points), and by temporal averaging over the entire post-virialization evolution to $\approx 9.4 \tau_{\rm ref}$. We note that within each of these 30 numerical simulations, $r_c$ and $r_t$ are uniquely obtained for the bimodal fit of Eq.~(\ref{eq:core-halo_fit}), as is $\tau_{\rm vir}$ from the corresponding decomposed energy evolution curves, and so the values  of all such quantities vary only slightly across different numerical realizations.

The consistency of the findings reported in this work is evident from these graphs. As expected, the scaled phase-space density, $\mathcal{D}/\mathcal{D}_\mathrm{ref}$, [Fig.~\ref{fig:A3}(a)] increases with decreasing radius from the soliton centre, reaching the critical value $\zeta(3/2) = 2.612 \sim O(1)$ approximately/just below the identified outer crossover radius $r_t$, and exceeding it by at least an order of magnitude at $r_c$. 

Moreover, Fig.~\ref{fig:A3}(b)-(c) demonstrate that the system still exhibits near-perfect coherence at the edge of the solitonic core, with $g_1(r_c) \approx g_2(r_c) \approx 1$, demonstrating a clear persistence of the suppression of both density and phase fluctuations within all such cores in our simulations. Coherence decreases as one moves radially outwards through the crossover region $r_c < r < r_t$, with the corresponding values at the outer edge of the crossover region reaching $g_1(r_t) \approx [0.6,\, 0.8]$ and $g_2(r_t) \approx [1.5,\, 1.8]$ (but $<2$ even when the non-negligible error bars are taken into account), with the higher total mass  configuration (150$M_{\rm ref}$) seemingly exhibiting slightly reduced coherence compared to the (50$M_{\rm ref}$) case. Nonetheless, these results clearly indicate that partial coherence still persists at radial distances at least as large as $r \approx r_t$. As expected, coherence is lost as one moves significantly beyond $r_t$, a feature demonstrated here through $g_1(10 r_c) \approx 0$ and the obtained mean numerical values satisfying $\bar{g_2}(10 r_c) \approx 2$ on average, albeit with the corresponding error bars placing its values in the range $[1.6,\, 2.4]$.

Finally, we note that in our work we have investigated the six largest eigenvalues of the single-particle density matrix, Eq.~(\ref{eq:PO}), and found out that for $n=[1,5]$ ($N_0=N_{_\mathrm{PO}}$) while most of the $N_{n}/N_{n+1}\sim 1$.

%---------------------------------------
%---------------------------------------



%\begin{thebibliography}{99}
%\end{thebibliography}

%%%%%%%%%%%%%%%%%%%%%%%%%%%%%%%%%%%%%%%%%%%%%%%%%%

%%%%%%%%%%%%%%%%% APPENDICES %%%%%%%%%%%%%%%%%%%%%
\clearpage
\newpage

% Don't change these lines
%\bsp	% typesetting comment
%\label{lastpage}
\end{document}